# Dumb-bell-shaped equilibrium figures

# for fiducial contact-binary asteroids and EKBOs

Pascal Descamps

Institut de Mécanique Céleste et de Calcul des Éphémérides, Observatoire de Paris, UMR 8028 CNRS, 77 av. Denfert-Rochereau 75014 Paris, France

Pages: 60

Tables: 2

Figures: 16

*Corresponding author:*
**Pascal Descamps**
IMCCE, Paris Observatory
77, avenue Denfert-Rochereau
F-75014 Paris
France

descamps@imcce.fr
Phone: +33 (0)140512268
Fax: +33 (0)146332834

**Abstract**

In this work, we investigate the equilibrium figures of a dumb-bell-shaped sequence with which we are still not well acquainted. Studies have shown that these elongated and nonconvex figures may realistically replace the classic “Roche binary approximation” for modeling putative peanut-shaped or contact binary asteroids. The best-fit dumb-bell shapes, combined with the known rotational period of the objects, provide estimates of the bulk density of these objects. This new class of mathematical figures has been successfully tested on the observed light curves of three noteworthy small bodies: main-belt asteroid 216 Kleopatra, Trojan asteroid 624 Hektor and Edgeworth-Kuiper-belt object 2001 $QG_{298}$. Using the direct observations of Kleopatra and Hektor obtained with high spatial resolution techniques and fitting the size of the dumb-bell-shaped solutions, we derived new physical characteristics in terms of equivalent radius, 62.5 ± 5 km and 92 ± 5 km, respectively, and bulk density, 4.4 ± 0.4 $gcm^{-3}$ and 2.43 ± 0.35 $gcm^{-3}$, respectively. In particular, the growing inadequacy of the radar shape model for interpreting any type of observations of Kleopatra (light curves, AO images, stellar occultations) in a satisfactory manner suggests that Kleopatra is more likely to be a dumb-bell-shaped object than a “dog-bone.”

## 1. Introduction

In the past decade, rotational light curves generated by unresolved close-binary systems or by single contact binary asteroids have been systematically modeled with two tightened Roche ellipsoids (Jewitt and Sheppard, 2002, Sheppard and Jewitt, 2004, Takahashi and Ip, 2004, Lacerda and Jewitt, 2007, Descamps et al., 2007a, Lacerda, 2008, Lacerda, 2011, Lacerda et al., 2014). These equilibrium figures of rotating liquid masses were introduced by Edouard Albert Roche (1820-1883) in 1847 to investigate the equilibrium shapes of a tidally locked binary system composed of a triaxial ellipsoidal body orbiting a rigid sphere along a circular Keplerian path. Later, Darwin (1906) extended the problem to a system composed of two mutually interacting fluid ellipsoids. Such a system could exist in equilibrium only for two limit values of the mass ratio $q$: $q = 1$ for congruent ellipsoids, and $q = 0$, if one of them is more massive than the other. More recently, nearly thirty years ago, Leone et al. (1984) revisited the problem for various mass ratios; for the sake of simplicity, they analytically calculated the ellipsoidal figure of each component while keeping the companion spherical. Strictly speaking, this is not the Roche problem but instead the "Roche binary approximation" as it was dubbed by the authors. Indeed, such solutions are pure approximations in the sense that tidal effects of a member's triaxiality on the companion's equilibrium shape are neglected. Moreover, another caveat arises from the assumption of a circular Keplerian orbit. Such an assumption is no longer tenable because the general dynamics of the orbit are not governed by a gravitational potential generated by a spherical mass.

The full and exact problem of equilibrium configurations of synchronous binaries has been numerically investigated only in the past few years (Gnat and Sari, 2010, Sharma, 2009). These investigations address the problem differently but reach the same conclusions: the "Roche binary approximation" seriously fails to reproduce the exact equilibrium solutions for small mutual separations. The equilibrium figures of tightly bound binaries are no longer triaxial ellipsoids, and departures from these pure ellipsoidal forms may amount to nearly 20%. The reason is that at closer separation, the bodies are both distorted by their mutual gravitational perturbations, and this distortion is most pronounced along their line of centers. However, the authors independently found that at mutual separations on the order of twice the sum of their mean radius, departures from ellipsoids given by the "Roche binary approximation" are negligible, and the latter can be successfully used for widely separated synchronous systems such as the most puzzling, 90 Antiope (Descamps et al., 2007b, Kryszczyńska et al., 2009, Descamps et al., 2009b, Descamps, 2010).

The purpose of this paper is to seek another set of figures able to be reliably substituted for Roche ellipsoids, which are somewhat inappropriate – and unrealistic – when dealing with the so-called contact binary asteroids. In this context, the term, and the model, "contact binary" was first coined by Cook (1971) to account for the unusually large amplitude of the asteroid 624 Hektor. Furthermore, as noted by Weidenschilling (1980), light-curve amplitudes above 0.9 magnitude cannot be produced by rotation of a single ellipsoidal figure, such as an ellipsoid of Jacobi. This is why models of asteroids with two ellipsoids in contact, or nearly in contact, were introduced to satisfactorily match such light curves. In fact, the term "contact binary" is now somewhat misleading. From a simple shape model made of two ellipsoids in contact, the "contact binary" concept gradually changed in terms of phenomenological interpretation. Currently, behind this designation is an implicit scenario of formation lending credence to the idea that a contact binary asteroid originates from low-velocity collisions that allow coalescence of a pair of asteroids into a single body without complete fragmentation. However, caution is needed here before making conclusions too quickly that such a scenario is the major, if not the unique, cause of subsequent formation of such puzzling bodies. Therefore, in the rest of the paper, the term "contact binary" only refers to a shape model of an asteroid with two ellipsoids in contact without any underlying assumption about its origin.

Furthermore, another feasible mechanism may exist that presents an opposing viewpoint, i.e., that such highly elongated bodies, with two prominent equal lobes at their ends, could instead be the outcome of an evolutionary process starting from a single body toward a well-detached binary system after undergoing a gradual shrinkage in the middle section. This is the fission hypothesis. In the transient stage, the figures are somewhat peanut-shaped and endowed with a high angular momentum. This family of equilibrium figures exists and can be described through the *dumb-bell equilibrium sequence* discovered and computed more than three decades ago (Eriguchi et al., 1982). This sequence branches off the Jacobi ellipsoid sequence from a bifurcation point that was noted early by Chandrasekhar (1966) as a point of secular instability. The equilibrium figures of this sequence are peanut-shaped or dumb-bell-shaped bodies composed of a bridge of material separating two equal lobes. This sequence ends by joining the congruent Roche ellipsoid family. Eriguchi and colleagues numerically computed a small sample of figures of the dumb-bell sequence, but this family of concave equilibrium shapes has been sidelined, mainly due to the impossibility of producing a tractable analytical description of such figures that is able to compete with the user-friendly Roche ellipsoids.

In the present work, the dumb-bell sequence is extensively and accurately computed. Each figure is fully characterized by its geometry and its physical properties. They are then compared with the classical Roche approximation as a toolkit used to fit light curves of putative contact binary asteroids. Fitting a light curve with the help of a dumb-bell-shaped model is equivalent to determining the value of a single parameter: the dimensionless angular velocity that uniquely identifies the dumb-bell figure. This parameter is a function of the rotational period and of the macroscopic density of the asteroid. In particular, we demonstrate that the dumb-bell model allows avoidance of some bias inherent to the Roche approximation in such a way that it is then possible to fully describe the asteroid not only in terms of overall shape and internal density but also in terms of light-scattering type by its surface without any preliminary, and consequently arbitrary, assumptions. Again, we stress that, in the rest of this article, the terms "dumb-bell-shaped" and "peanut-shaped" do not refer at all to any scenario of formation, whether fission or collision, but only to the best-matching figures derived for some peculiar asteroids. We remind the reader that the main interest in equilibrium shape models is to infer the bulk density of the object under consideration through the determination of the normalized angular velocity $\Omega$.

## 2. Computation of the dumb-bell equilibrium sequence

### 2.1. General problem

Let a fluid mass be rotating in space as a rigid body. Let us assume that it is isolated from other bodies and that its particles are subject to mutual attractions according to Newton's law of gravitation. We designate as the *z-axis* the axis of rotation of the mass and establish the origin of a rectangular system of axes *Oxyz* at the mass center. The condition fulfilled by an equilibrium figure of a homogeneous, incompressible rotating body is that, at any surface point, there is a perfect balance between self-gravity and the inertial acceleration due to the rotation. In terms of potential, the total potential, which is the sum of gravitational and rotational potentials, must be constant all over the surface. The equilibrium shape is thus an equipotential surface. Since Maclaurin discovered the axially symmetric figure of this equilibrium, many researchers have looked for non-axially symmetric equilibrium figures. It was Jacobi who obtained non-axially symmetric equilibrium figures for the first time: Jacobi ellipsoids. Since this discovery, much effort has been made to find a non-axially symmetric equilibrium with a non-ellipsoidal surface. Chandrasekhar (1966) found a neutral point on the Jacobi sequence. It is considered as a point where a secular instability sets in that is stimulated by the presence of viscosity in the fluid, resulting in a slow – or secular – departure from the unperturbed figure at a rate proportional to viscosity. Thus, in a slow evolution, the trajectory of an evolutionary figure would leave the Jacobi family at the point of bifurcation and subsequently follow a new family of non-ellipsoidal figures.

Chandrasekhar did not obtain any new sequence. He was mainly concerned with equilibrium stability. These figures are algebraic surfaces of the fourth order. Their existence was first proven by Eriguchi et al. (1982). We know that all figures of equilibrium, the existence of which have been demonstrated, possess at least one plane of symmetry. This plane is perpendicular to the axis of rotation. This is the x-y plane. Assuming another plane-symmetry about the x-z plane, Eriguchi and colleagues (1982) computed, for the first time, the equilibrium figures of a new sequence that branches off the Jacobi sequence at the bifurcation point found by Chandrasekhar. They demonstrated that the instability would therefore grow slowly with a furrow in the middle and gradually deepen to form non-axisymmetric dumb-bell-shaped figures.

To determine the equilibrium figures, we follow the approach earlier adopted by Eriguchi et al. (1982). The problem can be posed in the following way:

The radius vector of the free surface of an equilibrium figure can be expressed as

$$r = R(\mu, \varphi) \tag{1}$$

where $\varphi$ is the longitude, and $\mu = cos\theta$, $\theta$ is the colatitude.

A non-axisymmetric equilibrium is assumed, so that the following conditions must be satisfied that suppose the figure to be plane-symmetric about the x-y and x-z planes:

$$\begin{cases} R(\mu, \varphi) = R(-\mu, \varphi) \\ R(\mu, \varphi) = R(\mu, 2\pi - \varphi) \end{cases} \tag{2}$$

If the fluid mass is isolated in space, the boundary surface belongs to the set of surfaces of equal potential. For a given angular velocity, $\omega$, the equilibrium condition is that the total potential, i.e., the sum of the gravitational and rotational potentials, should have the same value everywhere on the surface. It is expressed as

$$U[R(\mu, \varphi), \mu, \varphi] = \Phi[R(\mu, \varphi), \mu, \varphi] - \frac{1}{2}R(\mu, \varphi)^2\omega^2(1 - \mu^2) = C \tag{3}$$

Here, $\Phi(r, \mu, \varphi)$ and $C$ are the gravitational potential and a constant, respectively. The problem should lead to the determination of a new equilibrium shape in such a way that Eq. [3] will be satisfied. Following the formalism of Eriguchi et al. (1982), the gravitational potential $\Phi(r, \mu, \varphi)$ of a homogeneous and incompressible fluid mass of density $\rho$ can be expanded as

$$\Phi(r,\mu,\varphi) = -G\int d^3r' \frac{\rho}{\left|r-r'\right|} = -\rho G\int_0^{2\pi} d\varphi'\int_0^{\pi}\sin\theta' d\theta' \int_0^{R(\mu',\varphi')} \frac{r'^2\,dr'}{\left|r-r'\right|} \qquad (4)$$

with $d^3r' = r'^2 \sin\theta' dr' d\theta' d\varphi'$ and $d\mu' = -\sin\theta' d\theta'$
and where G stands for the constant of gravitation

Adopting the legendre expansion,

$$\frac{1}{\left|r-r'\right|} = \begin{cases} \dfrac{1}{r}\sum_{n=0}^{\infty}\left(\dfrac{r'}{r}\right)^n P_n(\cos\gamma) & \text{for } r' \le r \\ \dfrac{1}{r'}\sum_{n=0}^{\infty}\left(\dfrac{r}{r'}\right)^n P_n(\cos\gamma) & \text{for } r' \ge r \end{cases}$$

where $\cos\gamma = \mu\mu' + \left(1-\mu^2\right)^{1/2}\left(1-\mu'^2\right)^{1/2}\cos\left(\varphi-\varphi'\right)$
then Equation $[4]$ can be rewritten as

$$\Phi(r,\mu,\varphi) = -G\rho\int_{-1}^{+1} d\mu'\int_0^{2\pi} d\varphi' \sum_{n=0}^{\infty} f_n\left[r, R(\mu',\varphi')\right] P_n(\cos\gamma) \qquad (5)$$

with the functions $f_n$ defined for any value of n>0 (with the exception of n=2) by:

$$f_n\left[r, R(\mu',\varphi')\right] = \begin{cases} \displaystyle\int_0^{R(\mu',\varphi')} \frac{r'^2}{r}\left(\frac{r'}{r}\right)^n dr' & \text{for } R(\mu',\varphi') \le r \\ \displaystyle\int_0^{r} \frac{r'^2}{r}\left(\frac{r'}{r}\right)^n dr' + \int_r^{R(\mu',\varphi')} \frac{r'^2}{r'}\left(\frac{r}{r'}\right)^n dr' & \text{for } R(\mu',\varphi') \ge r \end{cases} \qquad (6)$$

The functions $f_n$ eventually read

$$f_n\left[r, R(\mu',\varphi')\right] = \begin{cases} \dfrac{1}{n+3}\dfrac{R^{n+3}\left(\mu',\varphi'\right)}{r^{n+1}} & \text{for } R(\mu',\varphi') \leq r \\ -\dfrac{1}{n-2}\dfrac{r^n}{R^{n-2}\left(\mu',\varphi'\right)} + r^2\left(\dfrac{1}{n+3} + \dfrac{1}{n-2}\right) & \text{for } R(\mu',\varphi') \geq r,\ n \neq 2 \\ r^2\left[\dfrac{1}{5} + \ln\dfrac{R(\mu',\varphi')}{r}\right] & \text{for } R(\mu',\varphi') \geq r,\ n = 2 \end{cases} \tag{7}$$

Eriguchi et al. (1982) address the problem using the Newton-Raphson iterative scheme: let $S_0$ be the surface of a figure of equilibrium of a homogeneous liquid mass corresponding to the angular velocity $\omega_0$; $S_0$ is the initial guess, and it is the Jacobi ellipsoid at the bifurcation. After specifying a new value of $R(1,\varphi)/R(0,0)$ and linearizing Eq.[3] to first order in the small quantities $\delta R$, $\delta\omega$ and $\delta C$, they solve $R(\mu,\varphi)$, $\omega$ and $C$ using successive approximations. The solution is another surface $S_1$ found in the neighborhood of $S_0$ for a new value of angular velocity $\omega_1 = \omega_0 + \delta\omega$ differing, though only by a small amount from $\omega_0$. The solution $S_1$ is then taken as the initial guess for the next step corresponding to a new value of the ratio $R(1,\varphi)/R(0,0)$. Successive solutions along the sequence may thus be computed.

The main drawback is that the method fails to provide the solution for any value of $\omega$ given a priori because the choice of the initial guess should be very close to the solution. This problem can be circumvented by using a new class of surfaces, dubbed *cassinoids*, as relevant initial guesses. Cassinoids are concave figures built from a family of plane curves first introduced by Cassini more than three centuries ago as an alternative for Keplerian elliptic planetary orbits (see Appendix A). Interest in these figures slowly declined, but now they have unexpectedly found a second life in astronomy. These easy-to-use tridimensional bilobed figures can be mathematically handled through a unique shape parameter. Although cassinoids are axially symmetric about their main axis, with no flattening, they can be successfully applied as initial guesses, and the method converges toward an equilibrium solution after ~80 iterations with all the required precision ($10^{-4}$) according the following scheme:

Let us define the function $\Gamma$ by

$$\Gamma[R(\mu,\varphi),C] = \Phi[R(\mu,\varphi),\mu,\varphi] - \frac{1}{2}R(\mu,\varphi)^2\omega^2(1-\mu^2) - C \qquad (8)$$

For a given angular velocity $\omega$, we wish to solve $\Gamma = 0$ for the free surface $R(\mu,\varphi)$ and the constant $C$. From an initial guess $R_0$, which yields an initial value $C_0$ from Eq. [3], we can derive $R_1 = R_0 + \delta R_0$ with the Newton-Raphson iterative method. $R_1$ is the correction of the current guess by a small amount, $\delta R_0$, which is readily given by

$$\delta R_0 = -\frac{\Gamma[R_0,C_0]}{\frac{\partial\Gamma}{\partial R}(R_0,C_0)} \qquad (9)$$

where

$$\frac{\partial\Gamma}{\partial R} = \frac{\partial\Phi}{\partial R} - R\omega^2(1-\mu^2) \qquad (10)$$

with

$$\frac{\partial\Phi}{\partial R} = -G\rho\int_{-1}^{1} d\mu' \int_0^{2\pi} d\varphi' \sum_{n=0}^{\infty} \frac{\partial f_n}{\partial R} P_n(\cos\gamma) \qquad (11)$$

and

$$\frac{\partial f_n}{\partial R}\left[r, R(\mu',\varphi')\right] = \begin{cases} \dfrac{R^{n+2}\left(\mu',\varphi'\right)}{r^{n+1}} & \text{for } R(\mu',\varphi') \le r \\ \dfrac{r^n}{R^{n-1}\left(\mu',\varphi'\right)} & \text{for } R(\mu',\varphi') \ge r \end{cases} \tag{12}$$

The same procedure is then iteratively applied and repeated until $\delta C = C\text{-}C_0$ and $\delta R(\mu,\varphi)$ become negligibly small to within some numerical threshold fixed to $10^{-4}$.

## 2.2. Figures of the dumb-bell equilibrium sequence

Table 1: To be inserted

Each figure is modeled as a polyhedron made up of facets, whose vertices are defined over a square grid in longitude (φ) and latitude (θ), the sides of which are two degrees wide. The system of equations [8] is applied to N = 16471 vertex coordinates $R(\mu,\varphi)$. The two plane symmetries (Eq. [2]) restrict the problem to N/4 vertices distributed on the surface of a quarter-sphere. The geometrical and physical quantities of the solutions are summarized in Table 1. As a triaxial ellipsoid, we characterize the figure by five parameters: the semi-maximum dimensions along the three axes a, b, c – with a > b > c - and the waist polar semi-axis c' (the bilobed shape appears when $c' \le c$). Values of these geometrical quantities are given for an equivalent radius $R_e = 1$ (radius of a sphere of equal volume). The flattening and the elongation are defined, respectively, by the ratio *b/c* between the intermediate to short (polar) dimensions and the ratio *a/b* of the maximum equatorial dimensions. The dimensionless form of the angular velocity, the inertial momentum with respect to the axis of rotation and the angular momentum, $\Omega$, $\lambda$ and $H$, respectively, were used. These are defined as follows:

Fig. 1: To be inserted

The normalized angular velocity $\Omega$ (or normalized frequency) is defined by the ratio between the angular velocity and the critical spin rate for a spherical body $\omega_c$, which is the maximum spin rate that can be sustained by an undeformable body. At this spin rate, centrifugal forces would equal gravity at the equator of a spherical body.

$$\Omega = \frac{\omega}{\omega_c} = \frac{\omega}{\sqrt{\frac{4}{3}\pi\rho G}} \tag{13}$$

where $G$ the gravitational constant, and $\rho$ the bulk density.

The non-sphericity parameter $\lambda$ is defined as the ratio between the moment of inertia of the body with respect to its spin axis – considered as the maximum moment of inertia axis – and the moment of inertia of the equivalent sphere of radius $R_e$ (radius of the sphere of same volume)

$$\lambda = \frac{hMR_e^2}{\frac{2}{5}MR_e^2} = \frac{5}{2}h \tag{14}$$

where $M$ is the mass of the body, and $h$ is a coefficient depending on the shape (h=2/5 for a perfect sphere).

The specific angular momentum, $H$, is obtained by dividing the angular momentum by the following term, with the notation first introduced by Darwin (1887):

$$\sqrt{GM^3R_e} = MR_e^2\sqrt{\frac{4}{3}\pi\rho G} = MR_e^2\omega_c \tag{15}$$

With the adopted notations, the specific angular momentum can be conveniently written as

$$H = \frac{hMR_e^2\omega}{MR_e^2\omega_c} = \frac{2}{5}\lambda\Omega \tag{16}$$

The point of bifurcation found by Chandrasekhar corresponds, using our notation, to Ω = 0.3995 and H = 0.4841. This is the starting point of the new sequence. The sequence is plotted (crosses) in Fig. 1. Likewise, the congruent (mass ratio of the components equal to 1) synchronous binaries sequence computed by Gnat and Sari (2010) is shown (diamonds). For the sake of comparison and consistency, the filled circles denote the computations of dumb-bell figures first computed by Eriguchi et al. (1982). The solutions of the Roche approximation of a binary synchronous system made of twin ellipsoids (dashed line) are superimposed. These solutions drastically depart from the exact solutions (dashed line) at angular velocities greater than ~0.2. Moreover, no Roche solutions can exist beyond Ω ~ 0.325, and overlapping of ellipsoids occurs as soon as Ω is greater than 0.32.

Fig. 2: To be inserted

In Fig. 2, the figures of some typical models on the dumb-bell sequence are shown. Figures of equilibrium differ only slightly from Jacobi ellipsoids down to Ω ~ 0.38. From this point, the shape changes progressively to a dumb-bell-like shape, with two prominent lobes at each end. After that point, the waist of the dumb-bell becomes increasingly slimmer, and finally, the fission occurs as observed in Fig. 1, where the dumb-bell sequence smoothly and continuously joins the sequence of equilibrium configurations of tidally locked homogeneous congruent binaries. The overlapping figure (Ω = 0.2815 and H = 0.502) is, for the dumb-bell sequence, a figure where the furrow in the middle shrinks to nothing, and, for the congruent binaries sequence, two non-ellipsoidal tight bodies, nearly in contact (see Fig. 12 of Gnat and Sari, 2010). The solutions compare well: the flattening of one of the components of the congruent binary is 1.092 with an elongation of 1.43 (In normalized figure units, the full extent along X, Y, Z-axes are 2.52, 1.77 and 1.62. Gnat, 2014, private communication); they are respectively 1.097 and 1.43 for the corresponding dumb-bell solution. We can also note that the cassinoid corresponding to $e \equiv 1$ provides a very good approximated solution of this end-sequence equilibrium figure.

## 3. Application to some asteroids

### 3.1. Bilobed asteroids

To test the dumb-bell equilibrium figures, we focus, in this section, on three noteworthy bodies, each taken from the main asteroid groups, whose observational characteristics may qualify them for having strongly bifurcated shapes. Confirmation that elongated/bilobed contact binary asteroids, with a central concavity or waist and a modestly asymmetric shape, really exist noticeably came with the first estimations of asteroids' three-dimensional shapes from the radar images of near-Earth objects (Ostro et al., 2000, Benner et al., 2006). They must fulfill at least two inescapable salient conditions, typical of a putative dumb-bell-shaped body: a high degree of elongation, e.g., measured from radar or adaptive optics imagery, with the ratio of its maximum equatorial dimensions of ~2.4-2.8, and a high photometric range light curve (~ 1.2 mag which means that the brightness can change by as much as a factor of 3 during a full rotation) with U-shaped maxima and V-shaped minima of similar depth. However, the light curve minima become broader with increasing phase angle, which implies that V-shaped minima are highly suggestive of a contact binary asteroid only when viewed near zero phase angle (Lacerda and Jewitt, 2007). As far as the elongation is concerned, we can, based on the data in Table 1, state that the elongation of the dumb-bell-shaped figure varies from 2.8 to 3.6, which is substantially higher (~10-15%) than the most elongated observed asteroids. On the other hand, elongations lower than 3.0 are achieved for scaled angular velocities lower than ~0.30. As such, dumb-bell figures appear to be systematically slightly more elongated than the most elongated asteroids observed thus far.

For this point, we can rightly question the usefulness of fluid equilibrium figures to address real asteroid shapes as has recently been performed (Harris et al., 2009). Indeed, we must pay attention to this crucial point because, as noted earlier, density determination is a highly model-dependent result. It is evident that the theoretically possible figures are determined by a set of conditions that can concern either the physical state of the mass of a celestial body or its kinematical state. As for the physical state, the principal condition is that a celestial body is a fluid body. The fact that most asteroids are now considered as rubble piles, meaning that they are composed of solid boulders, as they appear to be, will not diminish the importance of results based on the fluidity hypothesis. Furthermore, some recent works tend to demonstrate that actual asteroid shapes are consistent with the evolution of aggregates tending toward minimum free energy states (Tanga et al., 2009a, Comito et al., 2011). These weak aggregates are held together by self-gravity. They seemingly have zero tensile strength, but this does not mean they are "fluid." However, they have the ability to re-shape if pushed to rapid rotation rates. Within this granular scheme, rubble-pile asteroids may exhibit many permissible spin and shape combinations (Holsapple, 2004). Yet Tanga and colleagues (2009a), from their numerical simulations, conclude that "when reaching a potential *valley*, the objects are very close in potential energy to both the Maclaurin and the Jacobi sequence. They thus can be stable and close to fluid equilibrium, despite being fairly different in shape." From subsequent simulations of the re-accumulation following a catastrophic disruption, Tanga et al. (2009b) demonstrate that natural shapes close to fluid equilibrium shapes are preferentially produced. They stress the importance of other non-disruptive shaping factors, such as minor impacts, tidal forces and seismic shaking, during the lifetime of rubble pile asteroids. In other words, it is not surprising not to find perfect equilibrium fluid figures because most of the solar system small bodies are not fluid but rather a collection of aggregates with some level of internal friction. Bodies' gravity tends to dominate over their internal strength, such that they gradually take the forms dictated by gravity and appropriate for their angular velocity. These forms, on the whole, are therefore well accounted for by equilibrium figures, but, locally, they depart from them with lumps and ridges that carve them out to produce their final bifurcated shapes.

Most of the time, we do not have observations other than those provided by photometry. A typical case of this is that first considered in the relation with the Edgeworth-Kuiper belt object 2001 $QG_{298}$. The two following test-asteroids, 216 Kleopatra – the famous "dog-bone-shaped" asteroid – and 624 Hektor, take advantage of high-resolution imagery (radar and/or adaptive optics), which undoubtedly have revealed their bilobed form. Moreover, they contain at least one satellite in orbit, which allows a direct measurement of their bulk density provided that their sizes are sufficiently well known, and consequently, their angular velocity $\Omega$. The asteroids 624 Hektor and 216 Kleopatra may thus be directly compared face-to-face to the dumb-bell model. Several light curve measurements are available for consideration for inclusion in this study: Sheppard and Jewitt (2004) and Lacerda (2011) for 2001 $QG_{298}$, Descamps et al. (2011) for Kleopatra and Dunlap and Gehrels (1969) for Hektor. The light curves are of suitable quality (RMS < 0.07 mag) for use in the analysis. All light curves are double peaked with a symmetrical shape, i.e., with both peaks and both minima having essentially the same brightness – a classic signature of a both elongated and bilobed body. Table 2 summarizes the results obtained for each object taken in consideration in the present paper in terms of angular velocity $\Omega$ and resulting bulk density $\rho$.

| Table 2: To be inserted |
|---|

### 3.2. General photometry of a dumb-bell-shaped figure

Theoretical light curve computations of polyhedral models were performed to compare their respective morphology and main features with the observed light curves as a function of the sole normalized rotational velocity $\Omega$. From a 3D shape model, once the line of sight and the direction of Sun are known, it suffices to select the facets that are both visible by the observer and illuminated by the Sun. The total reflected light is then computed by adding the contribution of each of these *active* facets. In the case where dumb-bells exhibit strong concavities, self-shadowing may occur at non-zero phase angles. This effect is taken into consideration because it represents a non-negligible amount of the overall brightness variation. Each facet reflects the solar light according to its orientation with respect to the Earth and Sun as well as according to the adopted scattering law. We adopted an empirical scattering model for the surfaces first proposed by Kaasalainen et al. (2001) combining, through a weight factor $k$, a lunar-type reflection – described by the Lommel-Seeliger law appropriate for low albedo rocky surfaces – and an icy-type law – Lambertian or diffuse reflection suitable for high-albedo surfaces with multiple scattering.

$$I = (1-k)\frac{\mu_0}{\mu_0+\mu} + k\mu_0 \qquad (17)$$

In the above equation, $\mu_0$ and $\mu$ are the cosines of the angles between the surface normal and the incidence and emission directions, respectively. A pure Lambert scattering is obtained with k = 1.0; this is the case of bright bodies with high limb darkening. Conversely, it is expected that a low-albedo body, with no multiple scattering and negligible limb darkening, has a small $k$ value. The sudden non-linear brightening toward small phase angles close to opposition – known as the *opposition surge* or *opposition effect* – is not taken into account. As the opposition effect occurs at very small phase angles for loosely packed regolith (phase angle $\alpha < 1°$) and most data are recorded at higher phase angles, light curves are weakly affected by the opposition effect. Because a dark surface would be dominated by single scattering, the *k* values are expected to be in the order of 0.1 (Kaasalainen et al., 2001).

Fig. 3: To be inserted

Illumination and viewing geometry is computed for each facet, and eventually, the total flux received by an Earth observer is simply the sum of the flux scattered by each visible facet. Figure 3 shows a set of light curves produced by the rotation of a dumb-bell object for two values of the angular velocity $\Omega$ (0.2815 and 0.32), the weighting parameter *k* of the scattering law being fixed to zero. The aspect is edge-on with zero phase angle. They are compared with the light curve generated by a pair of twin ellipsoids resulting from the Roche approximation in the case $\Omega = 0.32$, where Roche ellipsoids touch themselves, first for $k = 0$ (Fig. 3) and second for $k = 0.4$ adjusted to match the light curve amplitude produced by the rotation of the dumb-bell solution inferred for $\Omega = 0.32$ (Fig. 4).

Fig. 4: To be inserted

The light curve range varies between 1.0 mag and 1.4 mag. In addition to the large range of variability, light curves exhibit some general common features such that maxima are more smoothly rounded (U-shaped) than are sharp-edges minima (V-shaped). The *neck* between the two lobes plays a role in the overall shape of the light curve observed, forcing it to be more U-shaped than V-shaped. Several main trends can be drawn from these observations:

- Amplitudes become larger with increasing $\Omega$ values;

- At the same angular velocity and for the same weighting parameter, the depth of the two minima is always deeper with the dumb-bell-shaped figure. The gap can only be filled by the Roche model with a higher value of the angular velocity;
- For a given angular velocity, both models generate light curves of the same amplitude, provided that the weighting parameter used with the Roche model is significantly higher.

Fig. 5: To be inserted

By any standards, we see that using one model instead of another can eventually lead to quite different values of the normalized angular velocity – which is the key outcome expected from light curve fitting with equilibrium figures – or to unrealistic overestimated values of the weighting parameter implied by the Roche approximation. For visualization purposes, dumb-bell and Roche model images are displayed in Fig. 5 with their true rendering according to their respective scattering law. Shape models have the same equivalent radius and are computed for the same angular velocity $\Omega = 0.32$. The scattering parameter $k$ of the Roche model is adjusted to a value giving the same light curve amplitude (see Fig. 4). A significant limb darkening is visible on the Roche model ($k = 0.4$). Consequently, in the case where all other things are equal – angular velocity, aspect angle and light curve amplitude – the dumb-bell-shaped model is always less darkened towards the limb and slightly more elongated than its Roche model counterpart.

### 3.3. Amplitude-aspect method for shape and pole determination

In light curve fitting, we treat, as three independent parameters, the scaled angular velocity $\Omega$, the scattering parameter $k$ and the aspect angle $\psi$, which is the angle of then spin vector with respect to the line of sight ($\psi$ = 90° when the object is observed exactly equator-on and the aspect angle varies from 0° to 180°). As the amplitude of the light curve depends on these three parameters, we work out a specific method suitable for high-amplitude light curves with characteristic shapes to discriminate and quantify the effect of each of them. The method is analogous to the classical amplitude-aspect method (Zappalà, 1981, Pospieszalska-Surdej and Surdej, 1985), which is based on the assumed triaxial ellipsoidal model for the shape of the asteroid with a geometrical scattering law (in fact, it is supposed that the observed brightness of an asteroid is proportional to the instantaneous cross-section seen by a distant observer) and on the relationships between the aspect angle, the light curve amplitude and the asteroid magnitude at the light curve maximum, all obtained in at least three oppositions. These methods assume a triaxial ellipsoidal model for the shape of the asteroid. In the present method, the main difference is in the absence of any preliminary assumptions about the scattering properties of the asteroid surface and a shape model reduced to only one parameter, $\Omega$, instead of three (the three semi-axes a > b > c of the ellipsoid).

To this end, we look for a dumb-bell solution with the appropriate surface scattering and space orientation that minimizes the difference between the observed and synthetic light curve characterized by the following goodness-of-fit criterion $\chi^2$:

$$\chi^2 = \frac{1}{\sigma^2}\sum_i \left(L_{model,i} - L_{measure,i}\right)^2 \tag{18}$$

Where $\sigma$ describes the assumed common uncertainty in each point (typically ~0.02-0.05 mag), $L_{measure,i}$ is the value of the i$^{th}$ data point and $L_{model,i}$ the value of the modeled brightness for a given dumb-bell solution.

If we define the number of degrees of freedom ν = N-m, for *N* data points and *m* parameters (m = 3), we can introduce the reduced chi-square:

$$\chi_\nu^2 = \frac{\chi^2}{\nu} = \frac{RMS^2}{\sigma^2} \tag{19}$$

which provides a way to obtain the *root mean square error* or RMS error in our measurements. To find the best solution, in the sense of minimum $\chi^2$, we use a grid-searching method. We construct a library of light curves for comparison with observation through the entire parameter space studied. All possible grid points in the three-parameter space are searched. The limits of the parameter space are [0, 1] for k with a step of 0.1, [0.282, 0.34] for $\Omega$ with a step of 0.05 and [50°, 90°] for $\psi$ with a step of 0.5°. It appears that the best aspect angle, in the sense of the smallest $\chi^2$, while the weighting parameter *k* is held fixed, is roughly independent of $\Omega$ and *k*. The best aspect angle $\psi_{best}$ is thus taken as the mean value of $\psi$ for which confidence levels in the space parameter $\Omega$-*k* are then computed and plotted from the smallest value $\chi^2$ of the grid position.

To avoid the photometric effects due to the phase angle, light curves must come from observations made near opposition, for phase angles close to zero. To find the $1\sigma$ (corresponding to the 68.4% probability of finding the true values of the three parameters), $2\sigma$ (95.4%) and $3\sigma$ (99.7%) standard deviation regions encompassed by the joint variation of the three free parameters ($\psi$, $\Omega$, *k*), we draw the contour plots corresponding to the values of $\chi^2$ increased, respectively of 3.53 (solid line), 8.02 and 14.16 (dashed lines) from the $\chi^2$ minimum. As will be demonstrated, the best solution is not necessarily given by the smallest $\chi^2$. In fact, confidence levels isolate one or more solution regions in the space ($\Omega$, *k*) called "islets." In the case of several such regions, the best final solution can be univocally set out from at least two light curves by taking the overlapping islet. Reliable errors on the parameters are estimated from the full range of the outer limit of the largest overlapping $1\sigma$ islet.

Only two observations made at two different oppositions are needed to both solve for the shape solution and the scattering parameter, which are supposed to be identical in each case, whereas the aspect angle must preferably be changing to better constrain the final solution. Furthermore, the derived aspect angles allow inferring two opposite pole solutions. Indeed, we have the well-known relationship between the aspect angle $\psi$, the geocentric ecliptic longitude and latitude of the asteroid $\lambda_a$ $\beta_a$, and the ecliptic longitude and latitude of the spin axis $\lambda_p$ and $\beta_p$ (Zappalà, 1981):

$$\cos\psi = \sin\beta_a \sin\beta_p + \cos\beta_a \cos\beta_p \cos(\lambda_a - \lambda_p) \qquad (20)$$

The knowledge of the cosine of the aspect angle for a single observation means that all the possible rotational axes lie on a cone forming an angle $\psi$ with respect to the direction Earth-asteroid. At least two light curves obtained on different dates are needed to provide information about the changing aspect of the asteroid as it moves around its orbit. From these observations obtained, we may derive two pole solutions that define an identical orientation of the rotation axis. One, ($\lambda_{p0}$, $\beta_{p0}$), is the true pole. The other, ($\lambda_{p0}+\pi$, $-\beta_{p0}$), is a "spurious" pole that is the 180° opposite pole solution. For a low inclination orbit ($\beta_a$~0°), which is the most common case among asteroids, these two pole solutions equally satisfy Eq. [20], and the ambiguity in the pole solution cannot be eliminated from this method alone. This ambiguity is reflected in the value of the phase angle and, unless otherwise stated, the phase angle is always displayed between 0 and 90°. Several pairs of solutions, obtained in different oppositions, produce a set of pole coordinates, whose mean is adopted as the direction of the rotational axis. The standard deviation from the mean can provide a check of the obtained accuracy.

### 3.4. The Edgeworth-Kuiper-belt object 2001 $QG_{298}$

In 2003, the EKBO 2001 $QG_{298}$ was found to have an extremely large light variation of 1.14 ± 0.04 mag and a relatively long period of 13.7744 ± 0.0004 h (Sheppard and Jewitt, 2004). It was the first EKBO suspected of being a putative contact binary asteroid. Since then, the light curve obtained was successively fitted into the framework of the Roche binary approximation (Takahashi an Ip, 2004, Lacerda and Jewitt, 2007) and from the exact solutions of equilibrium configurations of tidally locked homogeneous binaries (Gnat and Sari, 2010). The scaled angular velocity $\Omega$ was derived, and, from the knowledge of the rotation period, the bulk density was determined to a value ranging from 0.59 to 0.72 g/cm$^3$ (Table 2). Seven years later, another light curve was measured showing a peak-to-peak photometric range of 0.7 ± 0.1 mag, which is significantly lower than in 2003 (Lacerda, 2011). This observed decrease in amplitude, caused by a change in viewing geometry, was interpreted by a large obliquity near 90 ± 30° with an aspect angle falling from $\psi$ = 90° in 2003 (equator-on) to nearly $\psi$ = 78° in 2010.

Fig. 6: To be inserted

These two light curves are reconsidered in the light of the present dumb-bell-shaped model. Owing to their high heliocentric distances, EKBOs can only be observed at small phase angles ($\alpha$ = 0.8° in August 2003 and $\alpha$ = 1.4° in August 2010). Figure 6 shows the confidence levels derived from each light curve. The contours were calculated by holding the aspect angle fixed at its optimum value - $\psi$ = 83.5 ± 1.0° in August 2003 and $\psi$ = 68.5 ± 1.0° in August 2010 while varying $\Omega$ and $k$. Leftmost plots show how the best dumb-bell model (solid curve) compares to the light curve data (points). Several *islets* appear to be permissible solutions. Although the $1\sigma$ contour is highly spread for the confidence levels related to the 2010 light curve, due to a poor RMS of 0.07 mag, there appears to be only one common islet that requires a unique solution given by the center of the islet: Ω = 0.318 ± 0.001 and k = 0.14 ± 0.03 (see Table 2). It stems from the fact the bulk density is 0.56 ± 0.02 g/cm$^3$. This is not a measurement of the density but just a model-dependent density, which is the result of a relevant model used to interpret the photometric data, and, as such, the associated uncertainty is *internal* uncertainty, which includes only measurement uncertainty and reliability of the applied shape model. The *external* uncertainty should include all input uncertainties, including for sidereal period, aspect angle, weighting parameter and, above all, shape model.

Our solution can be compared to the previous attempts to derive the physical properties. Takahashi and Ip (2004) could reproduce the observation only for a nearly equatorial view with an aspect angle of approximately 90°. Their best-fitted *k* value of the scattering parameter was k ~ 0.6-0.8. Smaller *k* values could not fit the observed light curve and amplitude too well. Likewise, Gnat and Sari (2010) cannot derive a solution with an angular velocity *Ω* greater than 0.282: the maximum value of their shape solutions. Consequently, they are forced to adopt a high value of the weighting parameter *k* to match the observed amplitude. Lacerda and Jewitt (2007), in their simulations, found that different surface properties – lunar or icy-type surface law – do not significantly change their density estimates. This is roughly true, albeit in Fig. 6, we can see that some preferred solutions, materialized in the form of islets, are well-identified for a set of discrete value of the weighting parameter. Lacerda and Jewitt adopted an equatorial-on geometry and found two solutions for, respectively, a lunar-type surface and an icy-type surface with no possibility of giving priority to one solution over another. In Fig. 7, we display the observed light curve (crosses) in 2003 along with the best-fit models coming from the dumb-bell solution (solid line) and the Roche solution (dashed line) found by Takahasi and Ip (2004) with a secondary to primary mass ratio q = 0.7. The RMS error is of 0.055 mag with the dumb-bell solution, whereas it is of 0.063 mag with the Roche solution. Thus, we may conclude that the dumb-bell model fits the data better than the Roche approximation. Furthermore, it leads to more realistic physical properties. The weighting parameter, for example, is in good agreement with what it is expected for a dark surface, k ~ 0.1 (Kaasalainen et al., 2001). In addition, there are no preliminary assumptions about the aspect angle, which is univocally determined by our amplitude-aspect method.

Fig. 7: To be inserted

From the derived aspect angles, we can infer the two J2000 ecliptic pole solutions equal to [110 ± 4°, -18 ± 17°] and [290 ± 4°, 18 ± 17°], corresponding to the respective aspect angles 83.5° in 2003 and 68° in 2010 or 96.5° in 2003 and 112° in 2010. The large uncertainties appear to confirm the large obliquity found by Lacerda (2011).

### 3.5. The main-belt asteroid 216 Kleopatra

The asteroid 216 Kleopatra has the largest amplitude of the large main-belt asteroids. Kleopatra was observed, for the first time in 1980 (Tholen, 1980), with an amplitude between 1.3 and 1.4 magnitude over the course of its 5.3853 ± 0.0003 h rotational period (Pilcher and Tholen, 1982). Twenty years later, early radar observations helped to develop the first polyhedral shape model of this M-type asteroid, which appeared as a double-lobed, narrow-waisted object (Ostro et al., 2000). In addition, recent adaptive optics imagery, conducted in connection with its unprecedented close opposition of September 2008, disclosed two small moons orbiting Kleopatra, which allowed the measurement of a bulk density of 3.6 ± 0.2 g/cm$^3$ (Descamps et al., 2011). As Kleopatra is classified as an M-type asteroid, this bulk density is suggestive of a highly unconsolidated rubble-pile internal structure if we adopt a metallic grain density of ~7 g/cm$^3$ (see discussion in Descamps et al., 2011). Its normalized angular velocity $\Omega$ was then determined at a value of 0.318 ± 0.045, which, once combined with a non-sphericity parameter of 3.65 inferred from its radar shape model, yields a specific angular momentum of 0.47. This immediately places Kleopatra near the dumb-bell equilibrium sequence (Descamps and Marchis, 2008a). Kleopatra is thus a worthy object to be tackled from the perspective of the dumb-bell-shaped model approach.

Fig. 8: To be inserted

From the observations collected during the photometric campaign carried out in 2008 (Descamps et al., 2011), the light curve taken on 23 September 2008, close to its opposition, was taken to derive the relevant dumb-bell-shaped solution. The phase angle was low enough ($\alpha = 8.2°$). From the two islets of confidence (Fig. 8), we retained the most realistic solution given by the smallest value of *k*, which is in agreement with the IRAS albedo of 0.12 confirmed by Takahsahi et al. (2004): $\Omega = 0.297 \pm 0.002$, $\rho = 4.23 \pm 0.11$ g/cm$^3$ and $k = 0.15 \pm 0.05$ (see Table 2) for an aspect angle of 79.5 ± 1° in agreement with the value (81.8°) given by the pole solution $\lambda = 76 \pm 3°$ and $\beta = 16 \pm 1°$ in J2000 ecliptic coordinates (Descamps et al., 2011). To check the validity of this solution, we applied it to the observation performed on 1 August 2008. At this time, the high phase angle ($\alpha = 19.93°$) was responsible for the observed high amplitude of ~1.3 mag because of the shadows cast successively by each lobe. This self-shadowing effect has been taken into account for deriving the synthetic light curves generated by each model. Figure 9 clearly shows that the radar shape model fails to satisfactorily reproduce the observed light curve, possibly due to the high asymmetry between its lobes, asymmetry which is not present in the observed minima. In contrast, the dumb-bell-shaped solution perfectly fits not only the observed amplitude but also the whole light curve. The intensity of the self-shadowing effect is maximum when the brightness is minimum and is rendered in Fig. 10. We have also computed the light curve from the other solution given by k = 0.41 and $\Omega$ = 0.288. The RMS error is 0.0294 mag, larger than the RMS of 0.0255 mag obtained with the previous solutions.

Fig. 9: To be inserted

Fig. 10: To be inserted

Figure 11 allows us to visually compare both models. The radar shape model of Kleopatra does not resemble the dumb-bell-shaped model in many respects. The elongations are quite different: 2.68 for the radar model and 3.17 for the dumb-bell model. The waist is most pronounced in the dumb-bell model. The lobes of the radar shape models are strongly asymmetric and have a more compact and compressed form. This done, the dumb-bell model can be directly compared and tested against two types of high resolution observation, on one hand with an adaptive optics (AO) image taken on 2009 December 7$^{th}$ with the 10-m Keck telescope (F. Marchis, private communication) and, on the other hand, a stellar occultation made a handful of days later, on December 24$^{th}$, 2009 (S. Preston and B. Timerson, private communication).

Fig. 11: To be inserted

Figure 12 shows a view of Kleopatra imaged in near-infrared with the 10-m Keck telescope in December 2009. A Laplacian filter was applied to highlight the edge of Kleopatra. The apparent aspect of Kleopatra, as seen on the plane of sky normal to the line of sight, was computed and displayed on the right side by using the dumb-bell model. Two extracted shape contours were overlaid on the AO image. Each of them corresponds to a given equivalent radius (radius of the sphere of the same volume) of Kleopatra. With a solid line, $R_e = 67.5 \pm 3$ km, the IRAS radius confirmed by the 2008 campaign (Descamps et al., 2011). With a dashed line, $R_e = 62.5 \pm 3$ km, the best-fitted value of the equivalent radius derived through a specific method described in Descamps et al. (2009a). This shortening of the equivalent radius comes from the longer form of the dumb-bell-shaped model (see Fig. 11). In passing, it emphasizes the high degree of dependence of the measured size on the adopted shape model. However, a better constraint on both shape and size can result from the observation of a stellar occultation involving Kleopatra a few days later, on 24 December 2009 at 11:59 UTC. This observing method is extremely powerful as a way of outlining very accurately the silhouette of an asteroid through its shadow cast on Earth during its transit in front of a bright star. In this case, the occulted star was TYC 4909-00873-1. The observing network was coordinated by the IOTA group (*International Occultation Timing Association*, http://www.occultations.org/) according to predictions made regularly by Dunham et al. (2013). Figure 13 displays the resulting occultation chords for the event derived by D. Herald with Occult 4.0 software (http://www.lunar-occultations.com/iota/occult4.htm). The aspect is basically the same as that of Fig. 12. In this case, the phase plays no part in the result. Among the observed chords, Kleopatra signals its presence with its fine-tapered silhouette, which seen from its longer side has a dumb-bell shape. Both shape models were fitted and applied to the silhouette. The best equivalent radius for the dumb-bell model was again adjusted to $62.5 \pm 3$ km. The dumb-bell outline is approximately 250 km long and 70 km wide, whereas the radar model outline is approximately 230 km long and slightly more than 80 km wide. The narrow waist of the silhouette is clearly visible and is better rendered by the dumb-bell model. Consequently, the AO image as well as the stellar occultation indicates that Kleopatra may be more of a dumb-bell than a dog-bone, which is not surprising as the overall uncertainty on the radar shape, as reported in Ostro et al. (2000), is large (20-25%). Ultimately, we see that Kleopatra appears to have a more lengthened and narrow form than generally believed through its radar model. The second point is that its equivalent radius should be revised to a smaller value of $62.5 \pm 5$ km. This pushes the bulk density, measured from the mass of Kleopatra derived from Kepler's third law (Descamps et al., 2011), to a value of 4.4 ± 0.4 g/cm$^3$, which is in excellent agreement with the value inferred from the angular velocity of the associated dumb-bell model (Table 2).

| Fig. 12: To be inserted |
|---|

| Fig. 13: To be inserted |
|---|

In addition to the radar model, several other shape models have been proposed. Tanga et al. (2001) derived from observations made with the HST/FGS interferometer a shape with two ellipsoidal components in contact. Adopting a lunar-type law, the overall size of their model is 273 ± 7 km x 75 ± 2 km x 51 ± 13 km. The large error for the minor axis c is due to the lack of sensitivity on this axis due to the nearly pole-on geometry of the asteroid during the observation. The inferred elongation (3.64 ± 0.20) is notably greater than the radar result but comparable to the elongation of the dumb-bell model, 3.2 (see Table 1). Their flattening is equal to 1.47 ± 0.36, which is also on the same order of the dumb-bell flattening of 1.1. With an equivalent radius of 62.5 km, the overall extension of the dumb-bell solution is 283 km x 89 km x 81 km, with a central waist of 38 km, which is in rough agreement with the results of Tanga et al. (2001). Takahashi et al. (2004b) focused on the question of Kleopatra's shape through the Roche approximation formalism. They obtained a Roche solution with a mass ration of 0.84 and a weighting parameter of 05-0.6 (Table 2). They notably reported that "we could not find a proper *k* value smaller than 0.1, and furthermore, with such small *k* values, no Roche binary models can explain all the observations." This makes it impossible of reconciling the inferred *k* value with the photometric properties of the dark surface of Kleopatra. This apparent incompatibility is eliminated through our dumb-bell-shaped solution. Eventually, both solutions have the same amplitude of 1.25 mag at 90° of aspect and 0° phase. This correspondence has already been raised in Fig. 4.

### 3.6. The Trojan asteroid 624 Hektor

Cook (1971) appears to have been the first to speculate that 624 Hektor was a close binary, but that was based on the high amplitude light curve. Cook's work followed Dunlap and Gehrels' discovery of the light curve's high amplitude and their correct identification of an elongated shape as the cause of the light curve inferred from the Hektor's rotational pole, which lies near the ecliptic (Dunlap and Gehrels, 1969). Depending on the orbital configuration, the light curve range varies between 0.1 and 1.2 mag with V-shaped minima and U-shaped maxima. At this time, most of the known highly elongated asteroids were among the smallest known asteroids – only a few kilometers in length – and believed to be plausible huge splinter-shaped fragments of a larger body. In contrast, Hektor, among the large asteroids, ranked third in the highest light curve amplitude. The fragment explanation was no longer ad hoc. Hartmann and Cruikshank (1978) made the first observations using simultaneous photometry and infrared (IR) bolometry to demonstrate that the maxima of the visual light curve correlates with maxima of IR thermal light curve, proving that the amplitude is caused by an elongated shape rather than being caused by albedo variations (which would produce anti-correlation). More than three decades later, in 2006, the final confirmation of the highly elongated/bilobed shape came from AO high-resolution imaging made at the Keck-II telescope (Marchis et al., 2006).

Fig. 13: To be inserted

We selected three light curves (Fig. 14) from the extensive dataset presented in Dunlap and Gehrels (1969) and used by Lacerda and Jewitt (2007) to fit a binary Roche solution. The differences in light curve range from one observation to another reflect the change of the observational geometry mainly through the aspect angle $\psi$. We obtain only one unique solution – or one common confidence islet – able to simultaneously best fit the observations at three observing campaigns (see Table 2). The weighting parameter, $k = 0.12 \pm 0.06$, indicates a dark lunar-type surface in agreement with a low albedo of ~0.06 (Fernández et al., 2003). Although the angular velocity determined by each of the solutions, Roche and dumb-bell model, is nearly the same, the light curve amplitude can be rendered with the Roche approximation only with a significantly higher value of the weighting parameter, $k = 0.5$. In fact, Lacerda and Jewitt made extensive use of a uniform scattering law – considered as equivalent to a lunar-type law with no limb darkening ($k = 0$) at small phase angles – but we were unable to replicate their results unless an increased $k$ value is used. They report a maximum amplitude of 1.2 mag at 90° of aspect and 0° phase, but we found it equal to 0.89 mag with $k = 0$ and 1.2 mag with $k = 0.5$. This moderate scattering agrees with its D-type and its red spectral slope characteristics of a rocky surface (Emery et al., 2011). Hektor is a member of the redder spectral group among Trojan asteroids. Patroclus, another Trojan asteroid, known as a nearly equal-sized binary with a density of $1.08 \pm 0.33$ $gcm^{-3}$ (Marchis et al., 2006b), is a member of the less-red spectral group resulting in a very different intrinsic composition. Using the three positions of Hektor in its orbit about the Sun and the three derived aspect angles (see Fig. 14), we can solve three systems of equations (Eq. [19]) for the pole solutions. There appears to be one common solution given in ecliptic J2000 coordinates: $\lambda = 324.6 \pm 0.9°$ and $\beta = -10.4 \pm 1.2°$, which is very different from the solution by Kaasalainen et al. (2002): $\lambda = 331°$ and $\beta = -32°$. It is not really surprising given the strong dependence of the pole direction on the shape solution.

Fig. 14: To be inserted

In the same way as for Kleopatra previously, the reliability of dumb-bell-shaped solutions derived for Hektor was tested on an AO image taken with the 10-m Keck telescope on 16 July 2006 at 13:50 UTC and published in Marchis et al. (2014). Figure 15 shows two overall contours superimposed on the AO image. Each contour corresponds to a value of the equivalent radius, 92 ± 5 km (dotted line) – the best fitted value - and 112 km (solid line). Marchis et al. (2014) derived a radius of 128 ± 3 km, which appears to be clearly overestimated. With such a size and from the new orbital parameters of the moonlet, they infer an extremely low bulk density of 1.0 ± 0.3 $gcm^{-3}$. There is a typo in the reported value of the semi-major axis in Marchis et al. (2014), and the right value is a = 957.5 ± 55.3 km (F. Vachier, private communication). With an orbital period P = 2.965079 ± 0.000288 days, the measured mass is therefore M = 7.914215 ± 1.408682 $10^{18}$ kg, which gives a bulk density of 1.32 ± 0.53 $gcm^{-3}$ for an object radius of 112.74 ± 17.86 km. With our new size of 92 km, which is 18% smaller, Hektor is approximately 416 x 131 x 120 km in size and the measured bulk density increases to the value 2.43 ± 0.35 $gcm^{-3}$, which is in very close agreement with the value derived from the knowledge of the scaled angular velocity (Table 2), 2.56 ± 0.05 $gcm^{-3}$.

## 4. Summary

Contact-binary asteroids with a bimodal appearance were suspected long ago based on their high photometric range light curves. The only tool available for modeling such unusual light curves relied on the figures drawn from the "Roche binary approximation" within a fully synchronized binary system, which is able to account both for the bilobed form and the distinguishable light curves. However, there is a serious caveat tied to their very conditions of use, i.e., when they are nearly in contact. In this case, the Roche binary approximation fails to properly represent the exact solution of the full Roche problem. However, the inferred solutions, albeit with these reservations, allowed the first coarse estimations of their bulk density. In the present work, this issue has also been tackled from another conceptual approach, taking advantage of the benefits provided by a new class of equilibrium figures: the dumb-bell-shaped figures. For the first time, we have reckoned and independently validated the dumb-bell sequence found by Eriguchi et al. (1982) in a more detailed and precise manner. In particular, we better assessed biases involved in the Roche approximation. Roche solutions lead to a systematic overestimate of the surface-scattering behavior (identified by the weighting parameter *k*) coupled with a smaller elongated shape. They are more often inconsistent with the global reflectance properties. Inferred results on surface properties should therefore be interpreted with caution and even not interpreted at all. Furthermore, quite obviously, the appearance of Roche solutions is clearly unrealistic, and, as such, they cannot be used to match resolved images or outline observations such as those provided by stellar occultations. Ultimately, solutions provided by the Roche approximation are useful only to the extent that they can be used as tractable models for fitting light curves and retrieving a first estimate of the bulk density and nothing else.

A more appropriate way to pursue the modeling of misshapen objects is through the bilobed models provided by the dumb-bell-shaped equilibrium figures. Dumb-bell-shaped figures of equilibrium (which, in the present case, could be dubbed *halteroids* from the French word "haltère" which is the translation of dumb-bell) proposed in this work are aimed at providing a new family of shapes parameterized by only one variable, the normalized angular velocity depending on the spin period and the bulk density. Using dumb-bell figures instead of Roche ellipsoids ultimately results in significant improvement of the overall consistency of the objects under study, as has been demonstrated in this paper : As regards the surface-scattering properties, dumb-bell solutions allow lower *k* values, which are more consistent with expectations, whereas the Roche solutions do not allow them; Dumb-bell-shaped figures are able to fit light curves remarkably well as well as other types of observations such as AO images or shape contour provide by stellar occultations. As a proof of reliability, inferred bulk density and surface scattering of the three fiducial bodies considered in this work, 2001 $QG_{298}$, Kleopatra and Hektor, fully agree with much of what has been obtained from other means of observation, clearly supporting the validity of the assumed fluid behavior.

We can also question the usefulness of using equilibrium figures in comparison with the powerful results achieved in the last ten years with the light curve inversion method (Kaasalainen and Torppa, 2001). Tridimensional shape models of asteroids can be reconstructed from the analysis of a set of light curves taken with various viewing and illumination geometries. However, the derived shape models are limited to global convex shapes. The difficulty is to tackle strongly nonconvex shapes, which can be dealt with only by using complementary multimodal data such as adaptive optics imagery, stellar occultations, interferometric HST/FGS observations, spacecraft flybys and thermal radiometry (Kaasalainen and Viikinkoski, 2012, Carry et al., 2012). These promising techniques are presently limited to a few targets, which should ideally have a large enough apparent diameter to be resolved and imaged beyond their star-like appearance. Only a few targets fulfill these conditions hitherto: 21 Lutetia (Carry et al., 2012), 22 Kalliope (Descamps et al., 2008b), 121 Hermione and 216 Kleopatra (Kaasalainen and Viikinkoski, 2012, Descamps et al., 2009). Kuiper belt objects are typical objects for which most of the available observations are nothing more than light curves. Accordingly, the equilibrium model remains the only solution, in the first instance, that is able to provide a first insight into an unknown distant object.

The figure of contact binary objects actually differs, slightly, from the figures given by the solutions of the problem of an isolated fluid mass in uniform rotation. Such a result opens up a new issue to be addressed: the dynamic reshaping under the action of various external effects, through a continuous adaptation of the shape to a new figure of equilibrium resulting from the displacement of the axis of rotation with respect to the axes of inertia.

## Appendix A: Cassinoids

The Cassinian ovals (or ovals of Cassini) were first proposed in the late seventeenth century by Giovanni Domenico Cassini (1625-1712) as a model for describing the movement of the Earth relative to the Sun. He rejected the Keplerian ellipses and believed instead that the Sun travelled around the Earth on one of these ovals, with the Earth at one of its foci.

| Fig. A1: To be inserted |
|---|

A Cassinian oval is a plane curve, locus of all points P such that the product of the distances of P from two fixed points has some fixed value, that is $\overline{PF_1}\,\overline{PF_2} = b^2$, where $F_1$and $F_2$ are two points located in (a, 0) and (-a, 0), respectively, and $b$ is a constant. Note the analogy with the definition of an ellipse (where product is replaced by sum). As for the ellipse, the two points $F_1$ and $F_2$ are called the *foci* of the oval. The shape of the curve depends on the ratio $e = {}^{b}/_{a}$. Furthermore, we will assume that $a < b < a\sqrt{2}$. The case $b =\geq a\sqrt{2}$ yields an oval, and the case $b = a$ gives another well-known curve, the Bernoullian lemniscate described by Jacob Bernoulli. The final case $a > b$ reduces to two disjoint ovals. The oval may be transposed to the tridimensional case after rotating this curve around its main axis joining the foci. The condition for a point P=(x, y, z) to lie on the oval becomes

$$\sqrt{(x-a)^2 + y^2 + z^2}.\sqrt{(x+a)^2 + y^2 + z^2} = b^2$$

As for the ellipse, the origin O is a symmetry center, and Ox and Oy are twofold symmetry axes. On squaring the two sides, we end up with the following quartic polynomial equation for the tridimensional surface of Cassini:

$$(x^2 + y^2 + z^2 + a^2)^2 = b^4 + 4a^2x^2$$

This surface is dubbed a “cassinoid” to echo “asteroid.” It can be expressed in a spherical coordinate system by substituting

$$x = R\sin\theta\cos\varphi$$

$$y = R\sin\theta\sin\varphi$$

$$z = R\cos\varphi$$

where θ and φ are, respectively, the colatitude and the longitude of the radius vector R (θ,φ). The parametric surface R (θ,φ) of a cassinoid eventually reads

$$R(\theta,\varphi)^2 = a^2\left[2cos^2\varphi sin^2\theta - 1 + \sqrt{(2cos^2\varphi sin^2\theta - 1)^2 + e^4 - 1}\right]$$

For convenience and without loss of generality, we may admit as unit length the half separation between foci so that a = 1. As long as $1 < e < \sqrt{2}$, the cassinoid describes a purely non-convex surface. As the parameter $e$ of the cassinoid gets smaller the region in the middle, near x = 0, gets thinner. Figure A1 shows the cassinoid solution for an arbitrary value of the parameter e = 1.04. The cassinoid has a longitudinal extension along its main axis of $2\sqrt{e^2 + 1}$ with a waist in the middle of $2\sqrt{e^2 - 1}$ in diameter; the two lobes have a maximum transversal size equal to $e^2$, which gives the general expression of the elongation of a cassinoid:

$$E_c = 2\sqrt{1 + e^2}/e^2$$

The maximum elongation is reached for e = 1 and amounts to $2\sqrt{2} = 2.83$.

**Aknowledgements**

The author wishes to thank Dr. Franck Marchis for providing him with adaptive optics images of Hektor and Kleopatra, and Dr. Frédéric Vachier for the orbital elements of the satellite of Hektor. The author is also grateful to Dr. Steve Preston and Brad Timerson of the International Occultation Timing Association (IOTA, http://www.occultations.org/) for providing him with the reduced stellar occultation by Kleopatra on December 2009.

**Table 1:** Geometrical and physical properties of equilibrium figures along the dumb-bell-shaped sequence. Semi-maximum dimensions along the three axes a, b, c – with a > b > c - and the waist polar semi-axis c' are listed. The quantities a, b, c and c' are expressed in units of equivalent radius.

| Ω | H | λ | a | b | c | c' | Flattening b/c | Elongation a/b |
|---|---|---|---|---|---|---|---|---|
| 0,3995 | 0,484 | 3,038 | 2,228 | 0,679 | 0,590 | 0,590 | 1,151 | 3,280 |
| 0,395 | 0,493 | 3,125 | 2,246 | 0,668 | 0,584 | 0,584 | 1,144 | 3,362 |
| 0,390 | 0,501 | 3,215 | 2,261 | 0,656 | 0,576 | 0,576 | 1,139 | 3,447 |
| 0,385 | 0,509 | 3,303 | 2,284 | 0,647 | 0,571 | 0,571 | 1,133 | 3,530 |
| 0,380 | 0,517 | 3,400 | 2,284 | 0,638 | 0,562 | 0,559 | 1,135 | 3,580 |
| 0,375 | 0,524 | 3,491 | 2,299 | 0,633 | 0,560 | 0,551 | 1,130 | 3,629 |
| 0,370 | 0,530 | 3,582 | 2,307 | 0,632 | 0,559 | 0,542 | 1,131 | 3,652 |
| 0,365 | 0,536 | 3,672 | 2,316 | 0,631 | 0,559 | 0,533 | 1,129 | 3,669 |
| 0,360 | 0,542 | 3,765 | 2,314 | 0,635 | 0,564 | 0,519 | 1,126 | 3,646 |
| 0,355 | 0,547 | 3,854 | 2,314 | 0,638 | 0,569 | 0,506 | 1,121 | 3,625 |
| 0,350 | 0,552 | 3,940 | 2,318 | 0,642 | 0,573 | 0,495 | 1,120 | 3,612 |
| 0,345 | 0,555 | 4,025 | 2,318 | 0,647 | 0,578 | 0,482 | 1,119 | 3,583 |
| 0,340 | 0,559 | 4,107 | 2,317 | 0,653 | 0,584 | 0,467 | 1,118 | 3,550 |
| 0,335 | 0,561 | 4,184 | 2,318 | 0,657 | 0,590 | 0,455 | 1,114 | 3,530 |
| 0,330 | 0,562 | 4,259 | 2,312 | 0,665 | 0,598 | 0,437 | 1,112 | 3,477 |
| 0,325 | 0,563 | 4,329 | 2,307 | 0,672 | 0,606 | 0,419 | 1,109 | 3,432 |
| 0,320 | 0,562 | 4,394 | 2,302 | 0,679 | 0,613 | 0,402 | 1,108 | 3,391 |
| 0,315 | 0,561 | 4,453 | 2,294 | 0,687 | 0,621 | 0,381 | 1,106 | 3,342 |
| 0,310 | 0,557 | 4,492 | 2,284 | 0,694 | 0,629 | 0,361 | 1,103 | 3,289 |
| 0,305 | 0,555 | 4,548 | 2,274 | 0,702 | 0,639 | 0,333 | 1,098 | 3,240 |
| 0,300 | 0,550 | 4,580 | 2,261 | 0,711 | 0,649 | 0,304 | 1,096 | 3,179 |
| 0,295 | 0,543 | 4,599 | 2,251 | 0,717 | 0,656 | 0,280 | 1,093 | 3,138 |
| 0,290 | 0,535 | 4,609 | 2,227 | 0,731 | 0,668 | 0,218 | 1,094 | 3,045 |
| 0,282 | 0,509 | 4,513 | 2,182 | 0,747 | 0,684 | 0,059 | 1,092 | 2,922 |
| 0,2815 | 0,502 | 4,409 | 2,152 | 0,754 | 0,687 | 0,005 | 1,097 | 2,854 |

**Table 2:** Physical parameters derived from several best-fit models.

| Solutions | Angular velocity (Ω) | density (g/cm³) | weighting parameter (k) |
|---|---|---|---|
| **KBO 2001 $QG_{298}$** | | | |
| Dumb-bell solution | 0.318 ± 0.003 | 0.56 ± 0.02 | 0.14 ± 0.03 |
| Roche solution (Takahashi and Ip, 2004) | 0.3 | 0.63 ± 0.20 | 0.7 ± 0.1 |
| Roche solution (Lacerda and Jewitt, 2007) | 0.312 | 0.59 ± 0.05 | 0 (lunar) |
| | 0.295 | 0.66 ± 0.05 | 1 (icy) |
| Congruent binaries (Gnat and Sari, 2010) | 0.282 ± 0.002 | 0.72 ± 0.04 | 1 (icy) |
| **624 Hektor** | | | |
| Dumb-bell solution | 0.298 ± 0.005 | 2.56 ± 0.08 | 0.12 ± 0.05 |
| Roche solution (Lacerda and Jewitt, 2007) | 0.302 | 2.48 ± 0,29 | 0.5[a] |
| **216 Kleopatra** | | | |
| Dumb-bell solution | 0.297 ± 0.004 | 4.23 ± 0.11 | 0.15 ± 0.05 |
| AO measurement & radar model (Descamps et al., 2009) | 0.318 ± 0.045 | 3.6 ± 0.4 | |
| Roche solution (Takahashi et al., 2004b) | 0.306 ± 0.003 | 4.0 ± 0,1 | 0.55 ± 0.05 |

[a] inferred from our own fit (see text)

# Figures

**Fig.1:** Dimensionless angular velocity *Ω* vs. dimensionless angular momentum *H* for the dumb-bell sequence (crosses). The Jacobi and the congruent synchronous binaries sequences are displayed as well. The physical properties of the synchronous binaries sequence shown with diamonds are taken from Gnat and Sari (2010). The filled circles denote the computations of some dumb-bell figures made by Eriguchi et al. (1982). Physical properties of each of these points refer to the values given in Table 1. The Roche approximation of synchronous twin ellipsoids (dashed line) departs drastically from the exact solution (diamonds) at angular velocities higher than 0.2. No Roche solution can exist beyond $\Omega \sim 0.325$. The transition between the dumb-bell sequence and the binaries sequence is smooth and continuous.

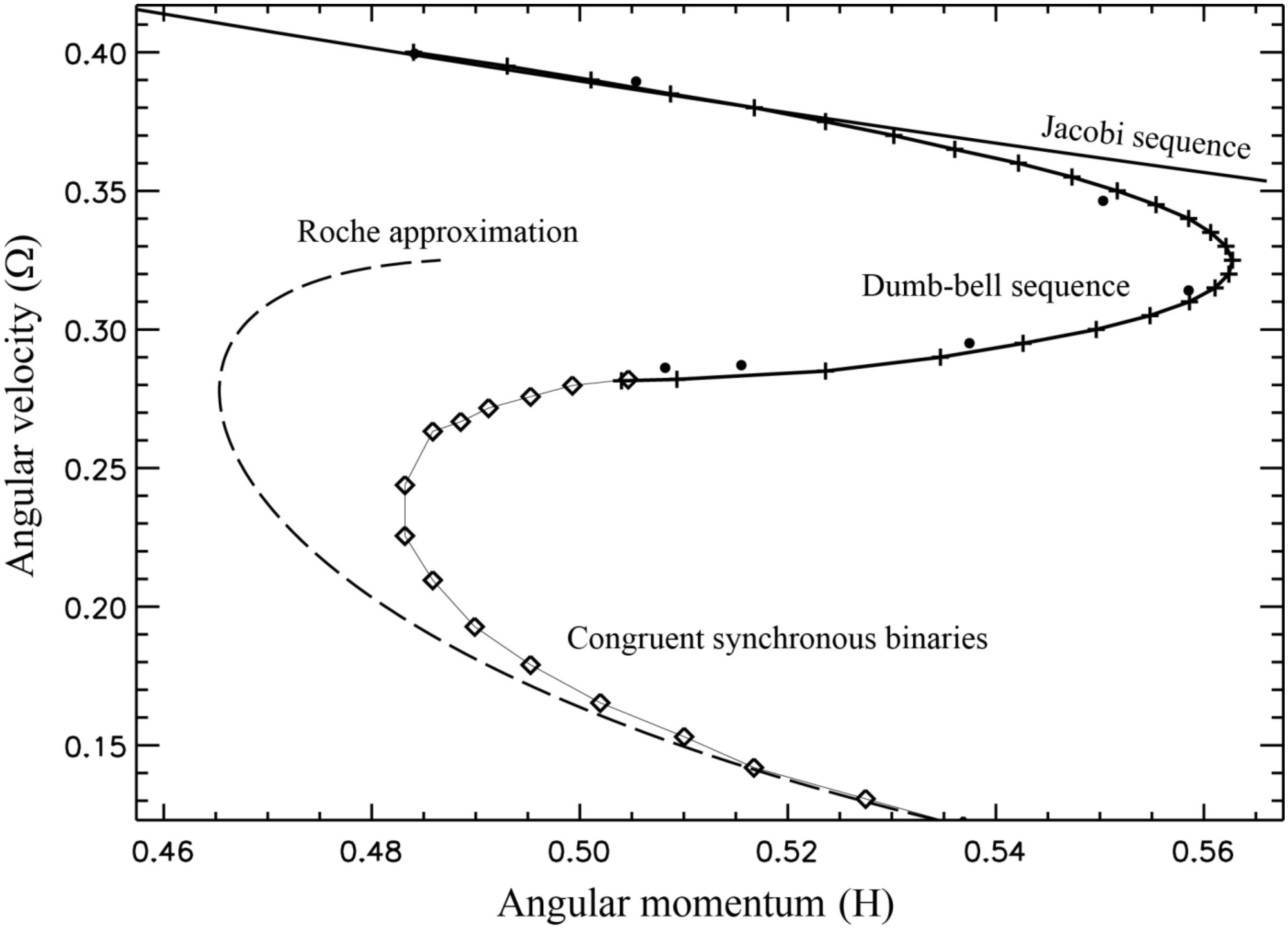

**Fig.2:** Sketch of various equilibrium shapes along the dumb-bell sequence for different values of the dimensionless angular velocity Ω. The left uppermost figure is the bifurcation Jacobi ellipsoid (at Ω = 0.3995, H = 0.484) where the Jacobi sequence becomes dynamically unstable. At this point, the dumb-bell shaped sequence branches off smoothly. From this departure ellipsoid, the middle portion of the dumb-bell shape narrows gradually as the angular velocity decreases. From $\Omega \cong 0.33$, two equal lobes appear at each end with a furrow in the middle of the figure which then looks like a peanut. The family terminates when the furrow shrinks to nothing (Ω = 0.2815, H = 0.502). The shape resembles to the starting equilibrium configuration of synchronous rotating bodies with two tight equal-mass components (see Fig. 12 of Gnat and Sari, 2010). The continuity between both families is clearly seen on Fig. 1.

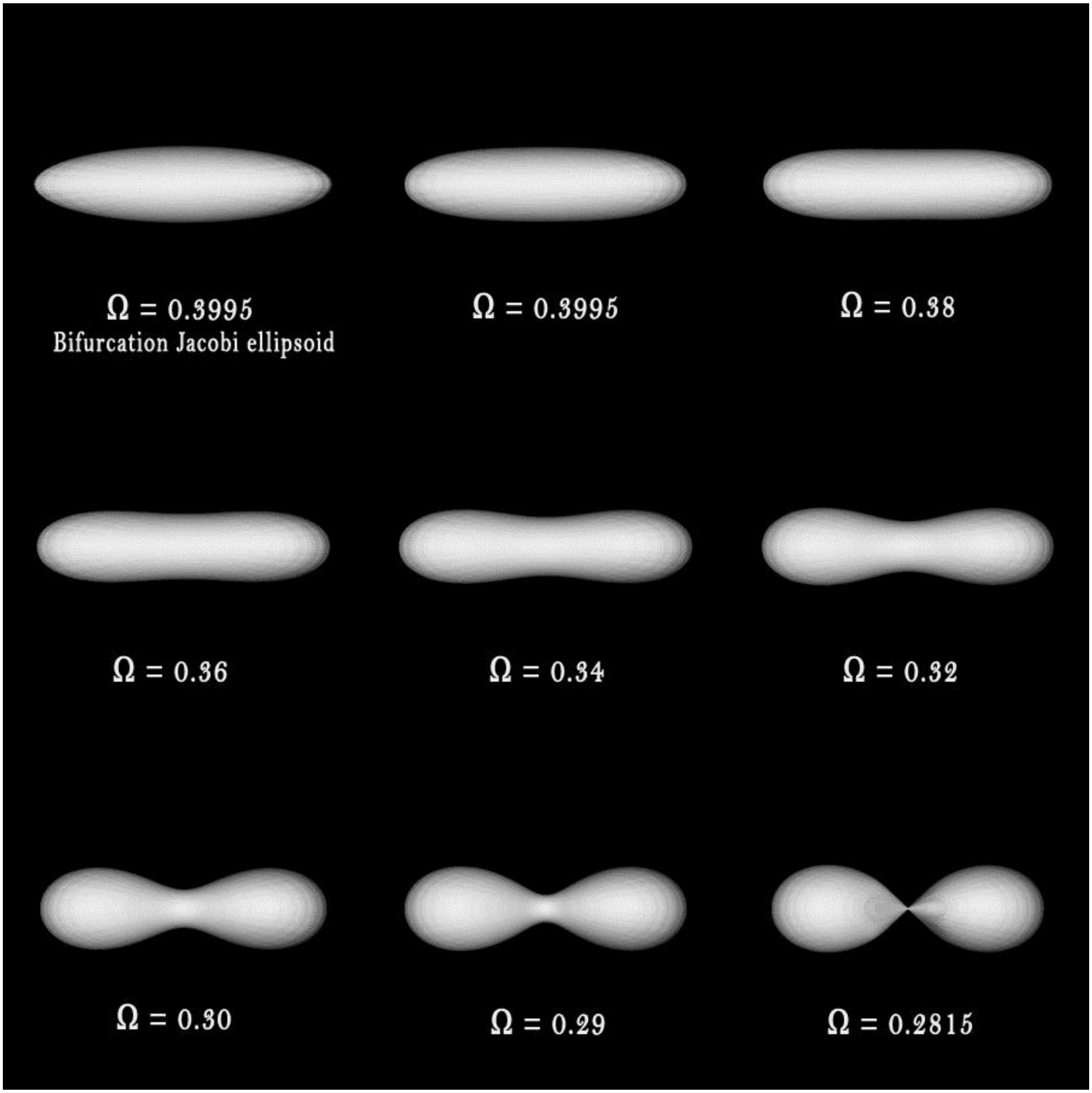

**Fig.3:** Light curves of a dumb-bell-shaped figure for different values of the scaled angular velocity $\Omega$ (0.2815, 0.32). The scattering parameter $k$ is fixed to 0, the phase angle is zero and the aspect angle is 90°. The light curve obtained with a pair of twin ellipsoids in the framework of the Roche approximation for $\Omega = 0.32$ (close to the limiting case) is superimposed (dashed line). Light curve amplitudes are greater than 1.0.

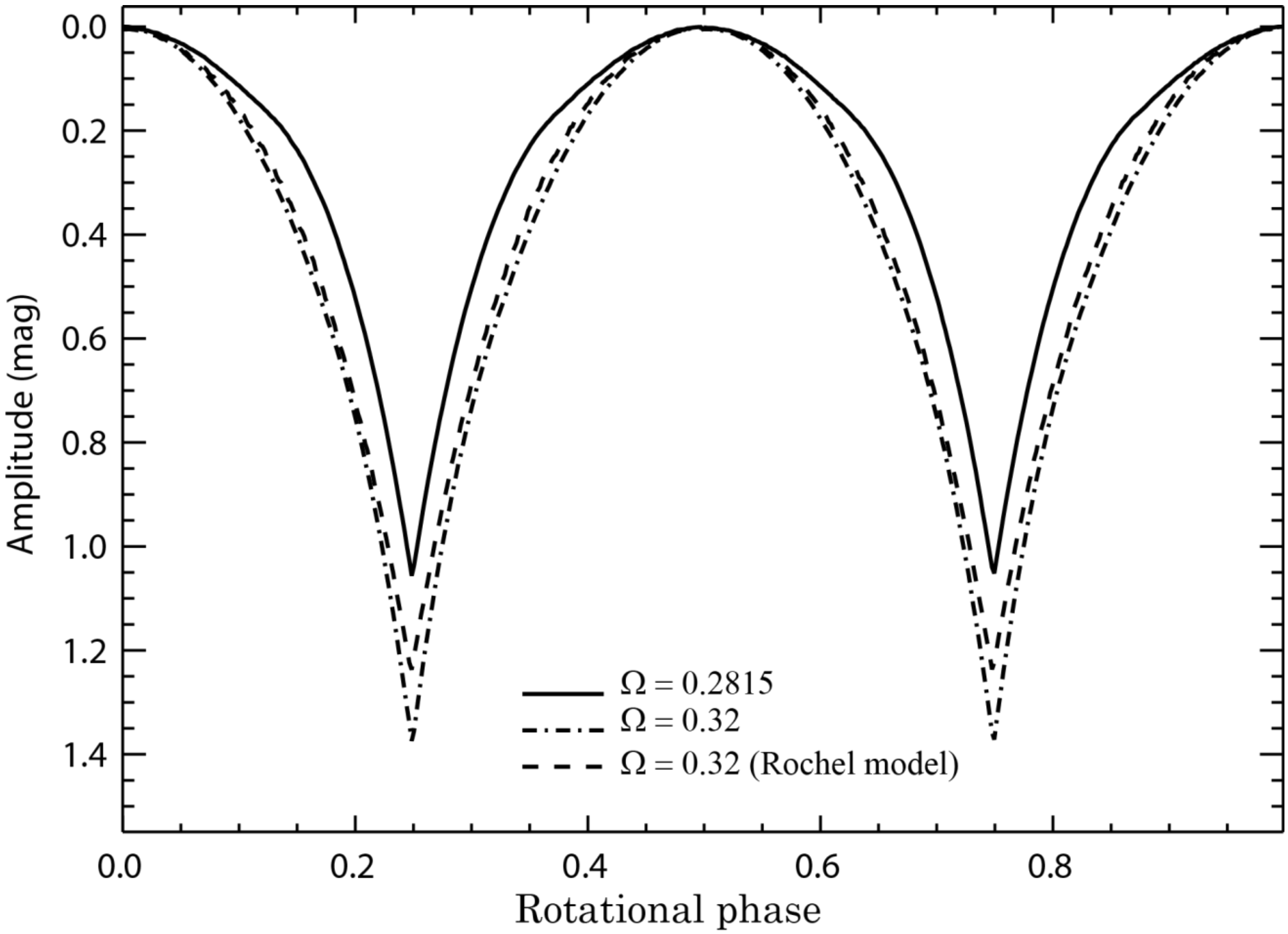

**Fig.4:** Comparison between light curves of equal amplitude derived for a same scaled angular velocity $\Omega$ = 0.32 at zero phase and for an aspect angle of 90°. In solid line, the light curve generated by the dumb-bell-shaped figure with k = 0. The dot-dashed line results from the Roche approximation adopting k = 0.4.

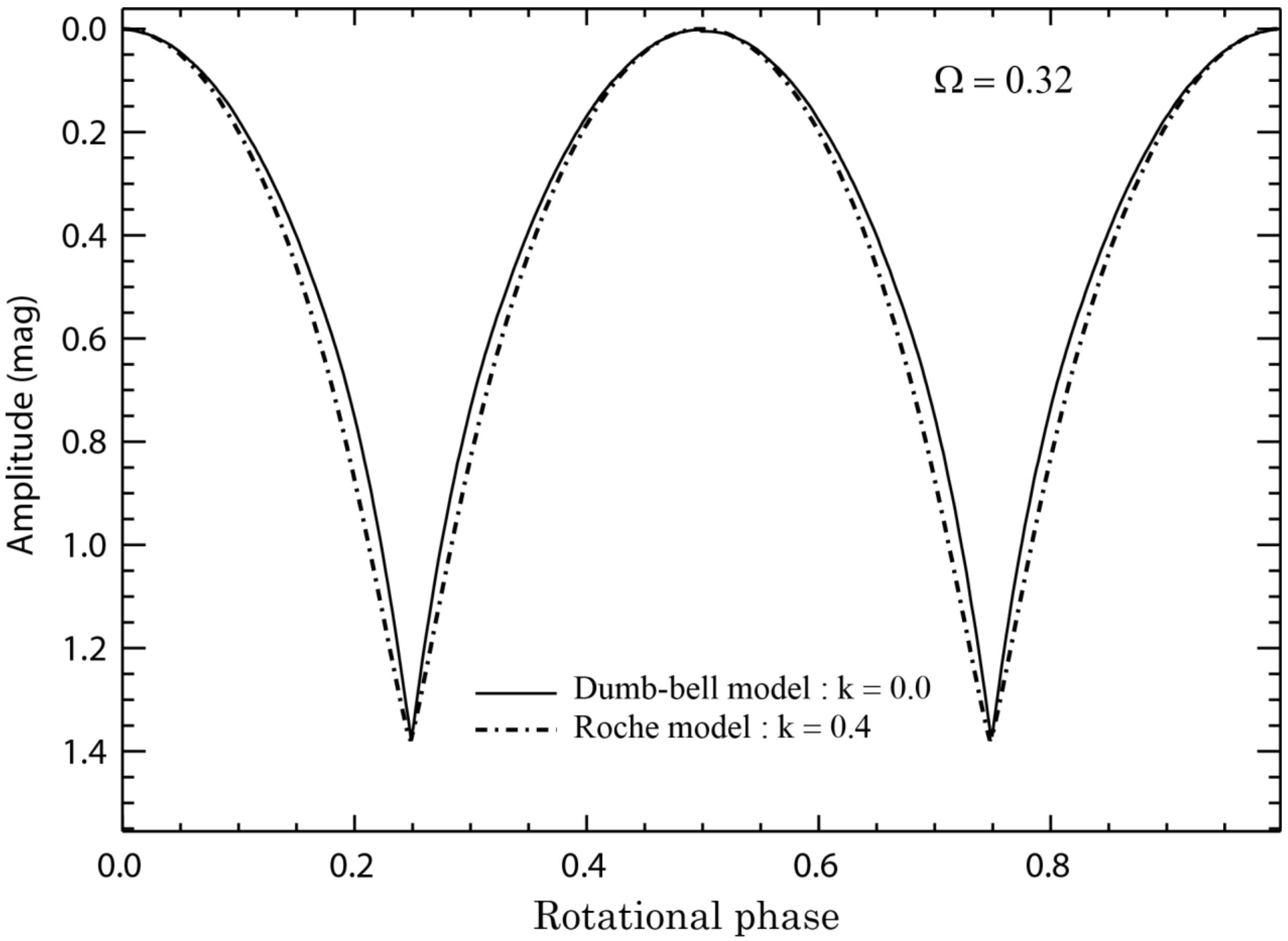

**Fig.5:** For visualization purposes, dumb-bell and Roche models images are displayed equatorial-on at opposition with their true rendering according their respective scattering law. Shape models have the same volume and are computed for the same angular velocity Ω = 0.32. The scattering parameter *k* of the Roche model is adjusted to a value giving the same light curve amplitude (see Fig. 4). A significant limb darkening is visible on the Roche model (k = 0.4). All other things being equal (amplitude, reduced velocity, aspect angle), the dumb-bell model is always slightly more elongated than its Roche counterpart.

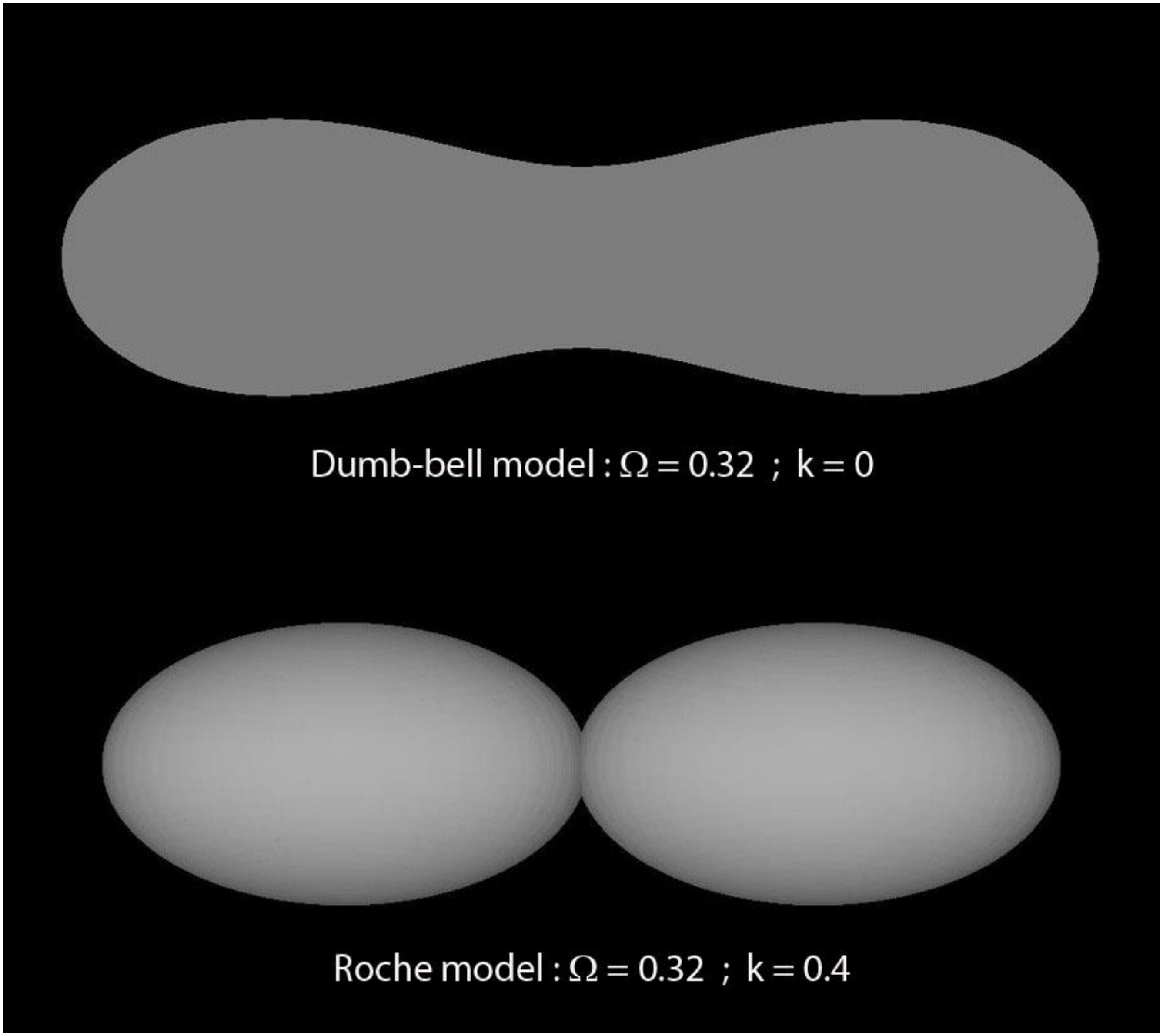

**Fig.6:** Set of confidence levels for the Kuiper belt object 2001 $QG_{298}$. The contours were calculated by holding the aspect angle fixed at its optimum value while varying $\Omega$ and $k$. The small contours, drawn with solid lines, give the region of the Ω-k space in which there is a 68.4% probability of finding the true values of the two parameters (1σ region). The large contours, shown as dashes lines, correspond to the regions 2σ (95.4%) and 3σ (99.7%). Observed light curve for the Kuiper belt object 2001 $QG_{298}$ (data points) and best fit solution (overlapping 1σ region in solid line) is obtained for $\Omega = 0.318 \pm 0.001$ and $k = 0.14 \pm 0.03$. With an assumed uncertainty of 0.05 mag, this solution yields in 2003 $\chi^2 = 187.7$ for $\nu = 167$ degrees of freedom which gives a reduced $\chi^2$ of 1.12; in 2010, $\chi^2 = 112.1$ for $\nu = 60$ degrees of freedom which gives a reduced $\chi^2$ of 1.86.

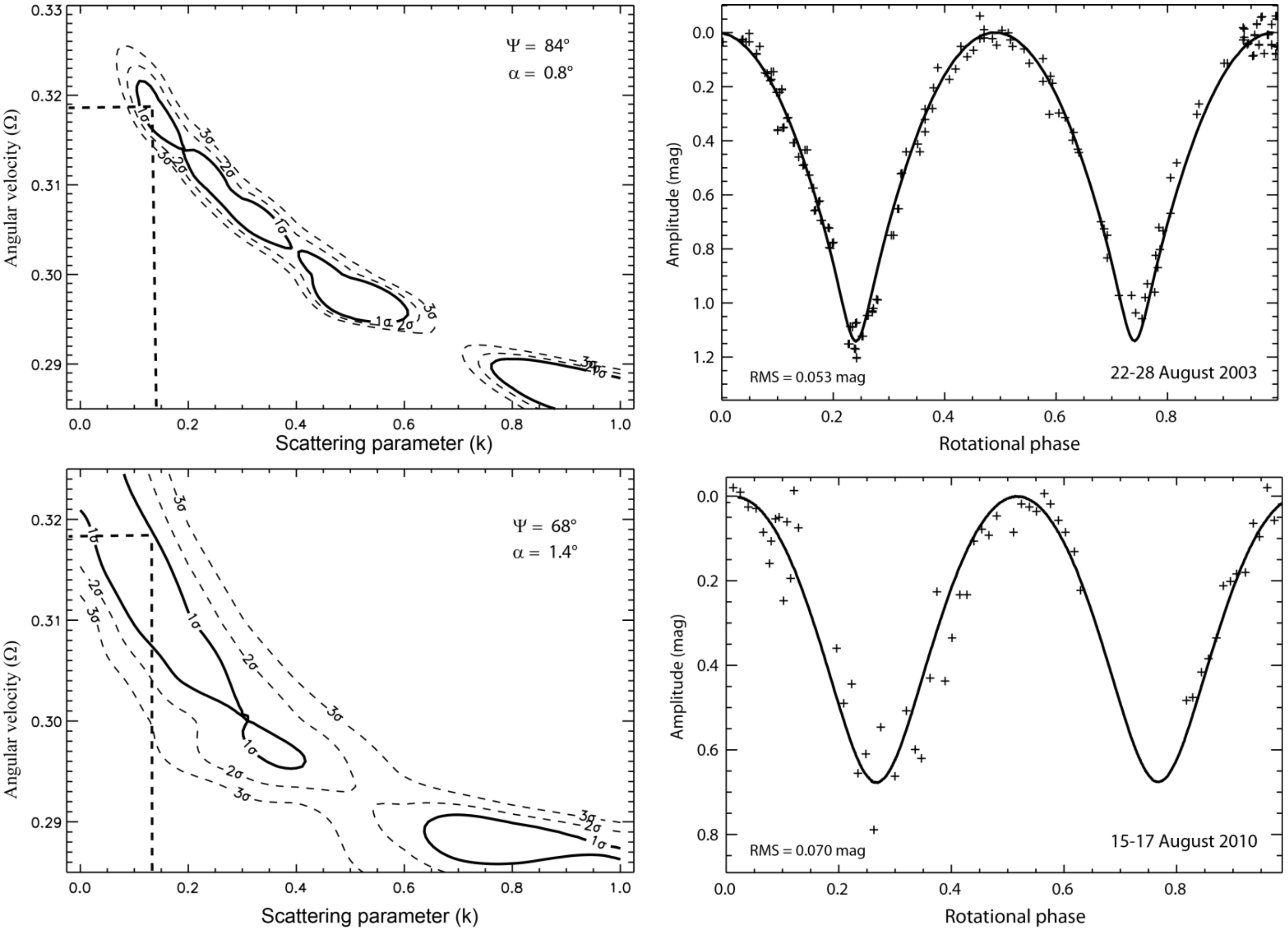

**Fig.7:** Equatorial view of the best shape models of 2001 QG298 according to the dumb-bell or the Roche approach. The Roche solution is taken from Takahashi and Ip (2004) with a secondary to primary mass ratio q = 0.7. Resulting light curves have been superimposed to the 2003 observation: in solid line, the light curve caused by the dumb-bell model, and in dashed line the light curve caused by the Roche solution. The RMS error is σ = 0.053 mag for the dumb-bell model and 0.063 mag for the Roche model.

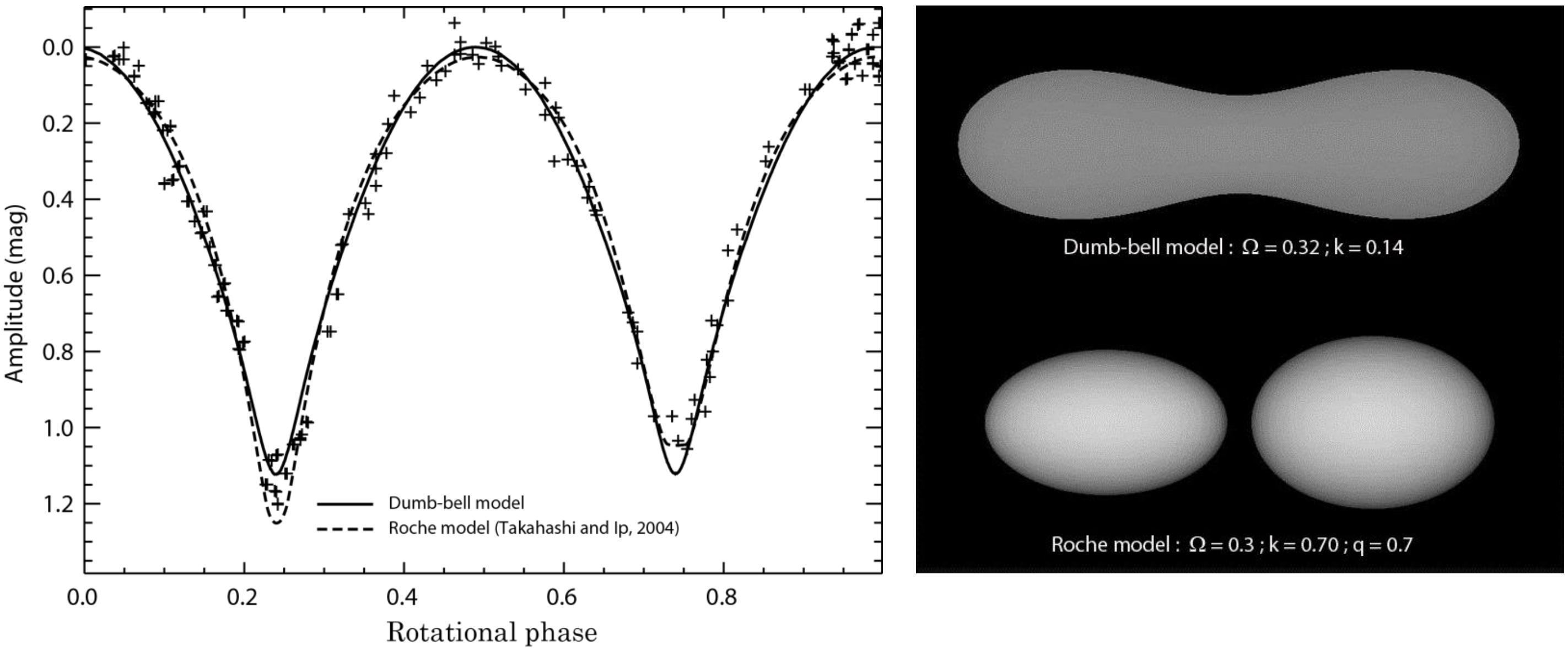

**Fig. 8:** Set of confidence regions for 216 Kleopatra derived from light curve taken on 23 September 2008. See legend of Fig. 6 for the general description of the contours. Two possible solutions appear. The solution $\Omega = 0.298 \pm 0.002$ and $k = 0.14 \pm 0.05$ is preferred because it gives a moderately low scattering parameter. With an assumed uncertainty of 0.025 mag, this solution yields $\chi^2 = 446.5$ for $\nu = 446$ degrees of freedom which gives a reduced $\chi^2$ of 1.00.

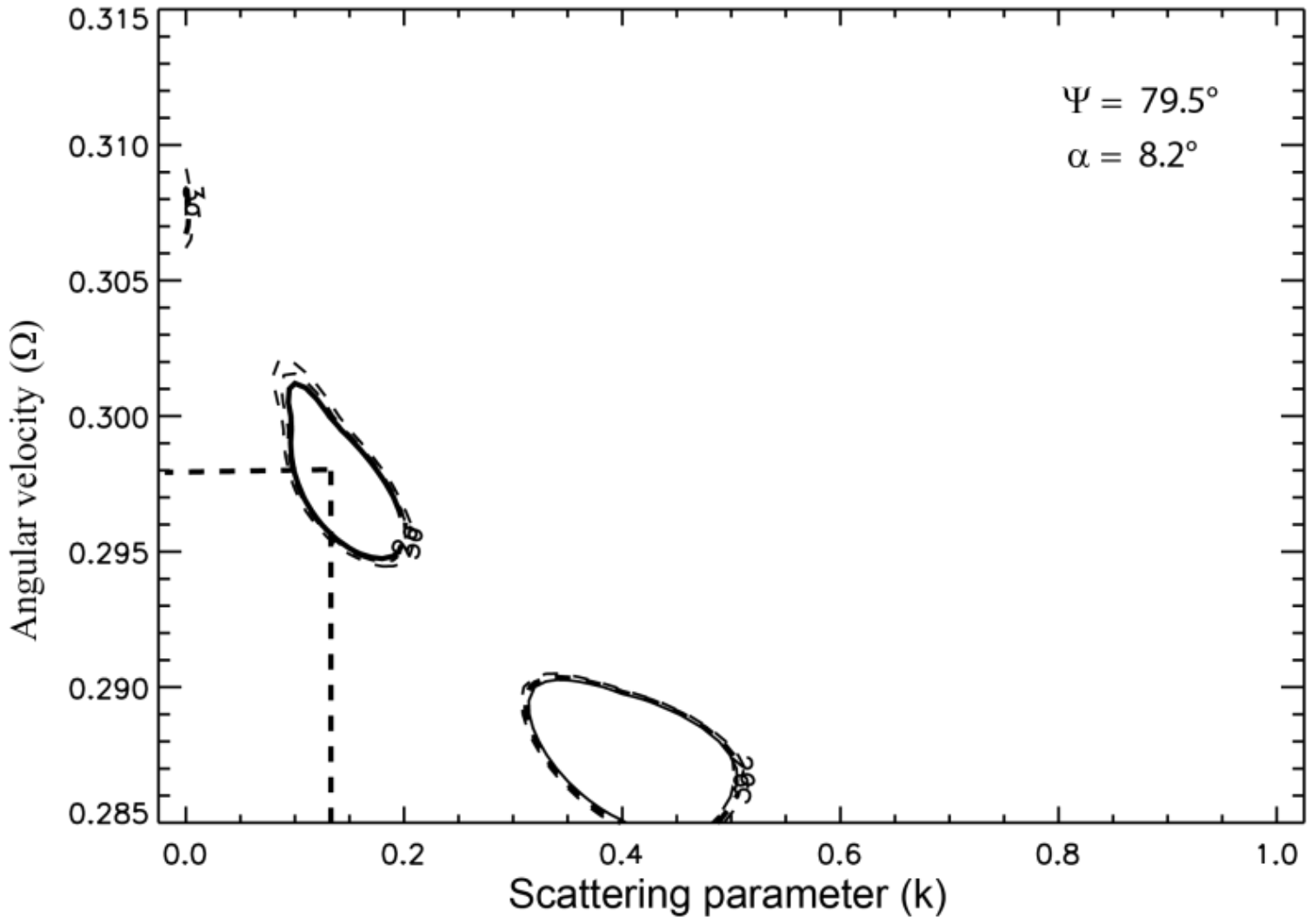

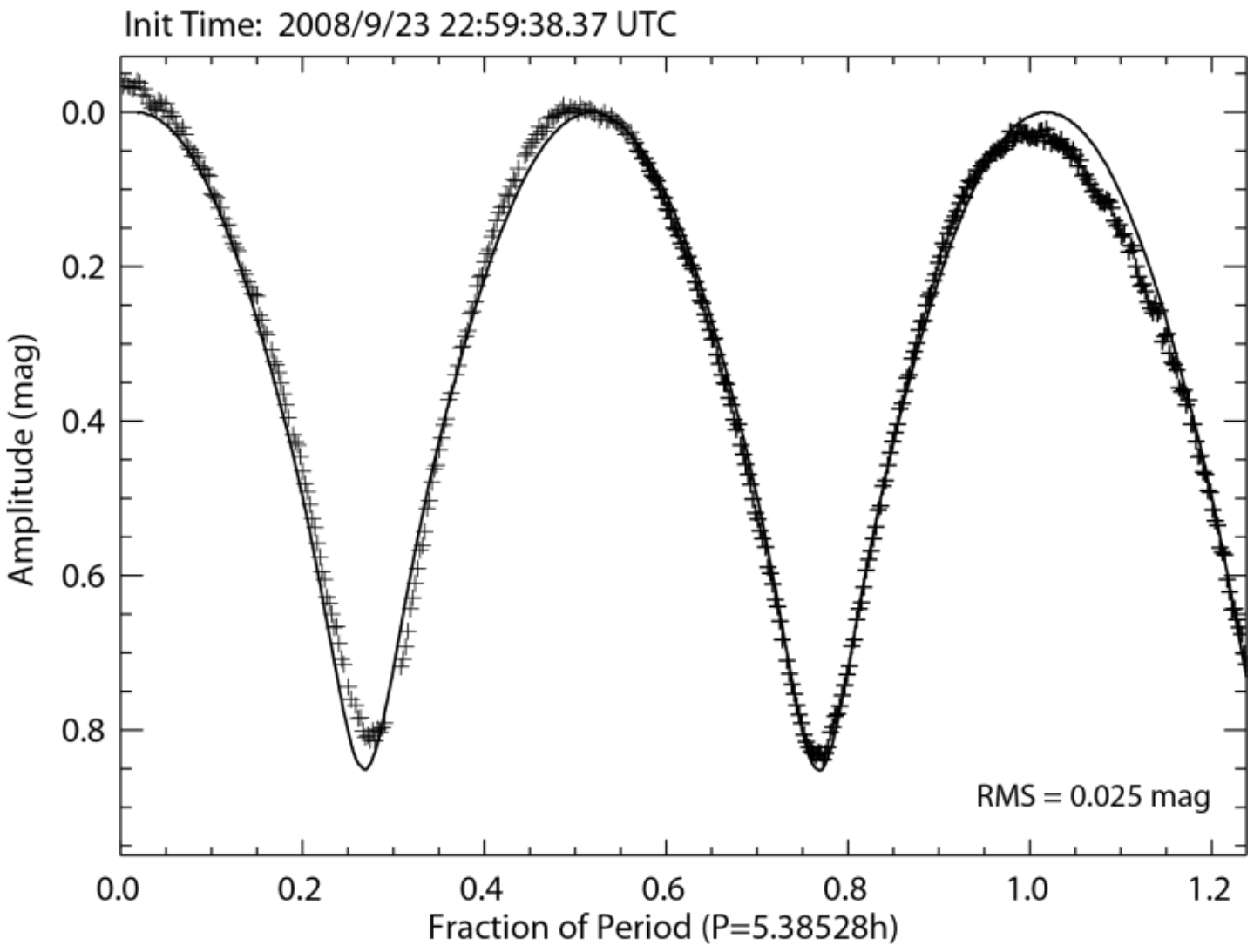

**Fig. 9:** Observed light curve for Kleopatra (data points) taken on 1 August 2008. The best fit solution (solid curve) was obtained with a dumb-bell model for Ω = 0.298 and k = 0.15. The dashed curve comes from the dog-bone radar model with the same scattering law applied to the surface. The high amplitude of ~ 1.3 mag – rarely observed – is due to a huge self-shadowing effect yielded by the lobes at a high phase angle of 19.9° (see Fig. 10). The radar shape model obviously fails to satisfactorily reproduce the observed light curve, possibly owing to the high asymmetry between both lobes.

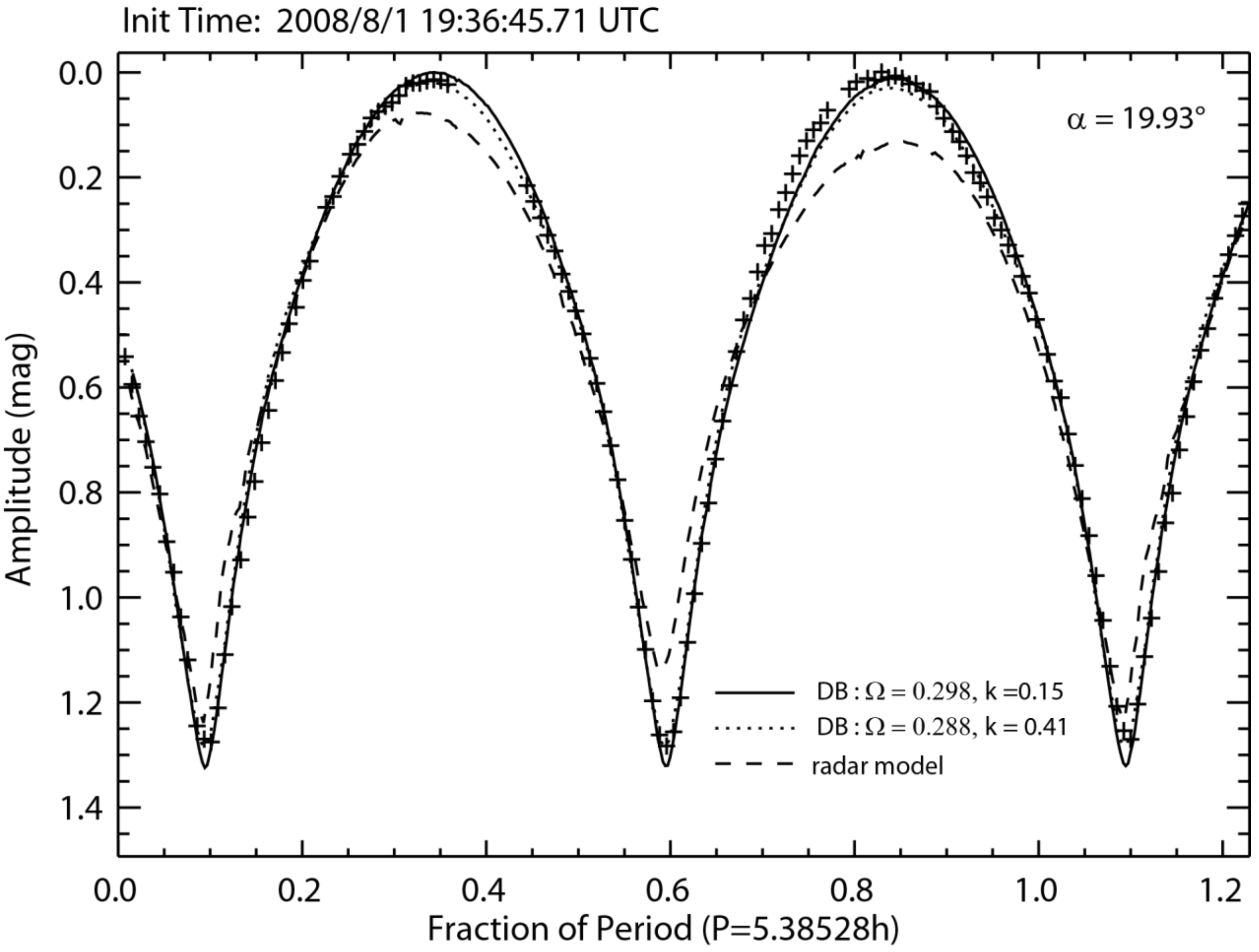

**Fig. 10:** Effect of self-shadowing on both models (dumb-bell and radar) on 1 August 2008.

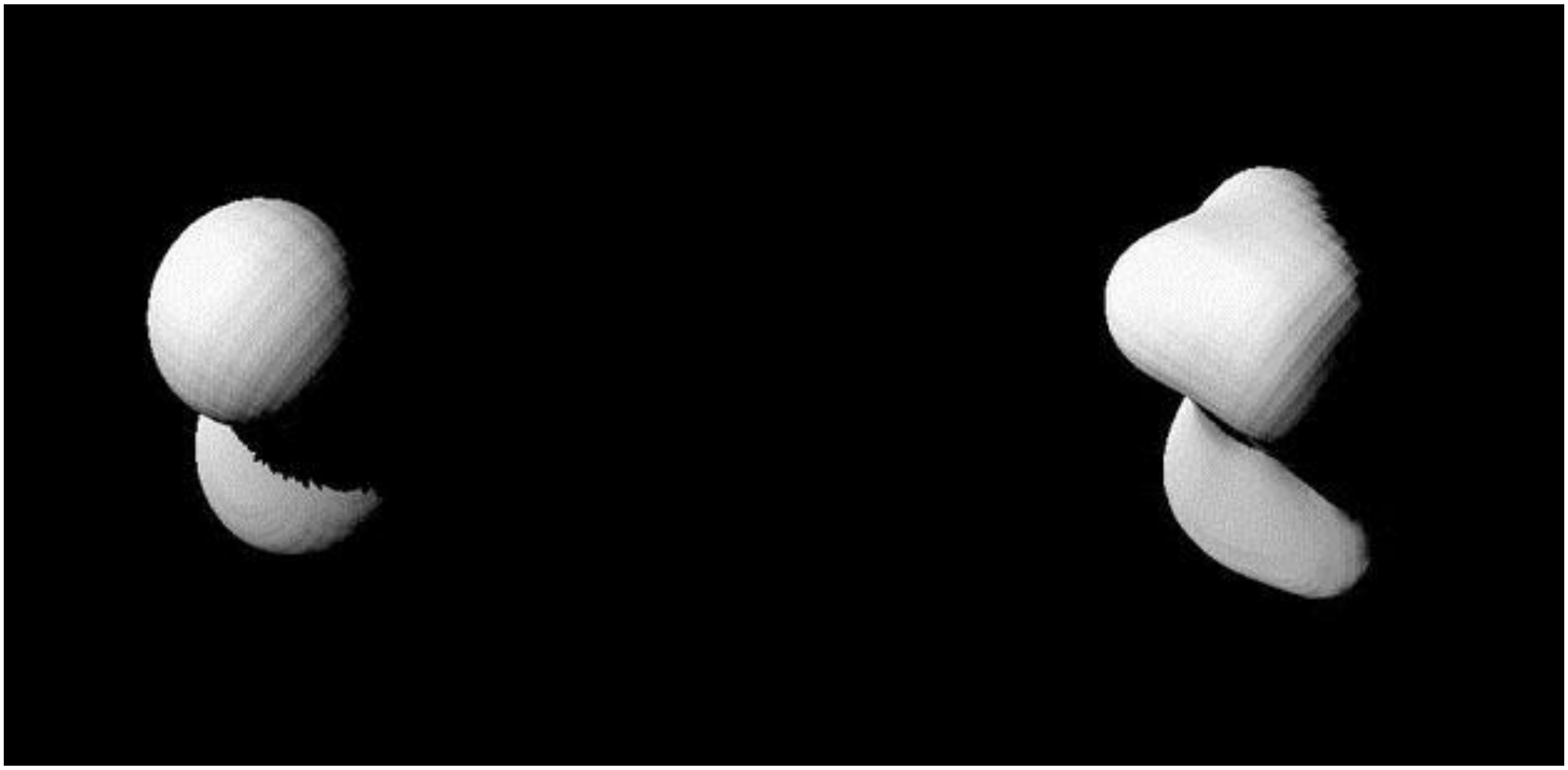

**Fig. 11:** The “dog-bone” radar model and the dumb-bell-shaped model of Kleopatra displayed at the same scale.

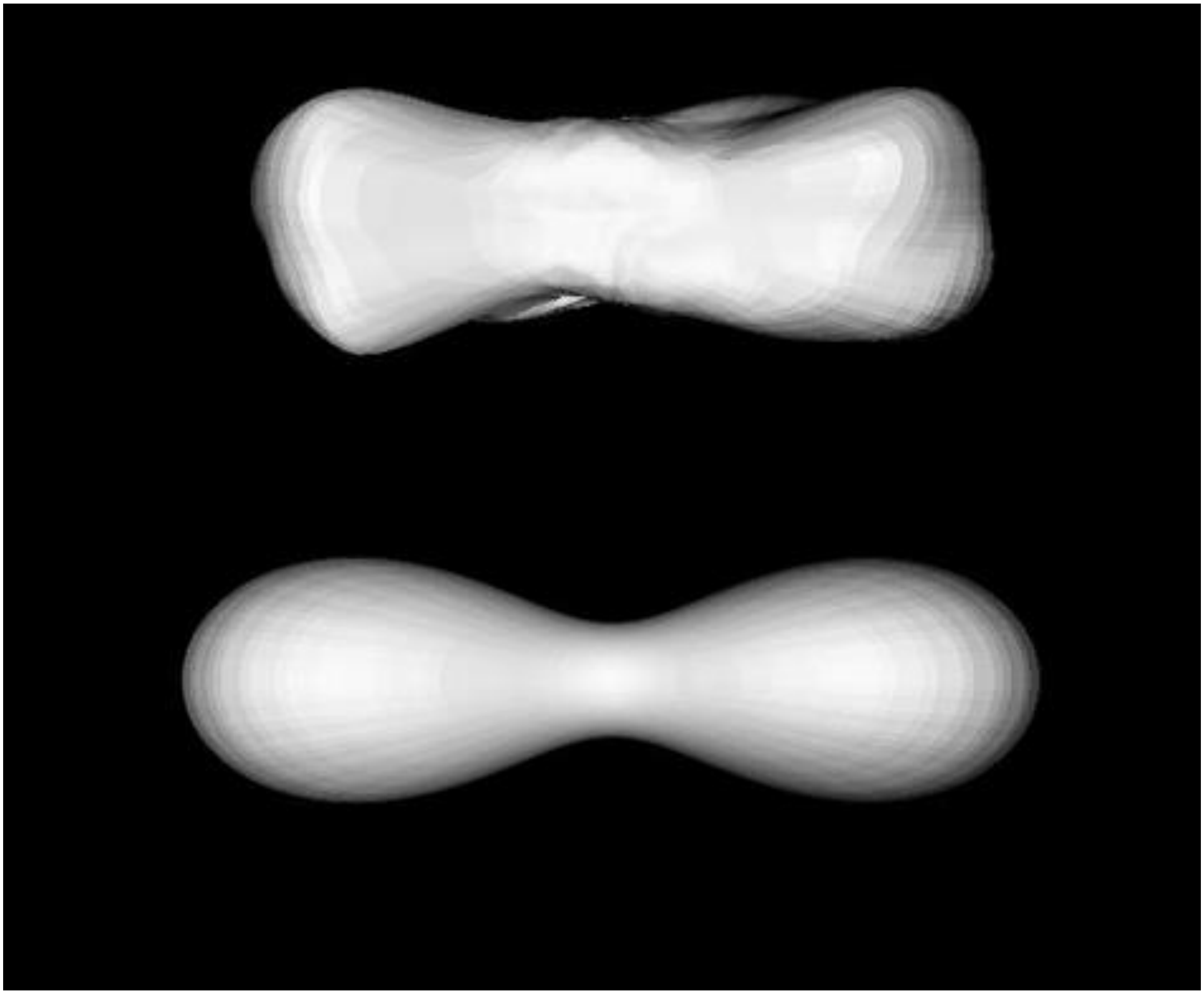

**Fig. 12:** Near-infrared AO image of Kleopatra taken with the 10-m Keck telescope on 2009 December 7th at 16:06 UTC. At this epoch, a scale of 100 mas represents 169 km. The phase angle is large (20.9°). North is up and east is to the left. A laplacian filter was applied in order to highlight the edge of Kleopatra. Two extracted contours of the dumb-bell-shaped model, each corresponding to a value of the equivalent radius ($R_e$ = 67.5 km in solid line, and $R_e$ = 62.5 km in dashed line), were superimposed to the AO image.

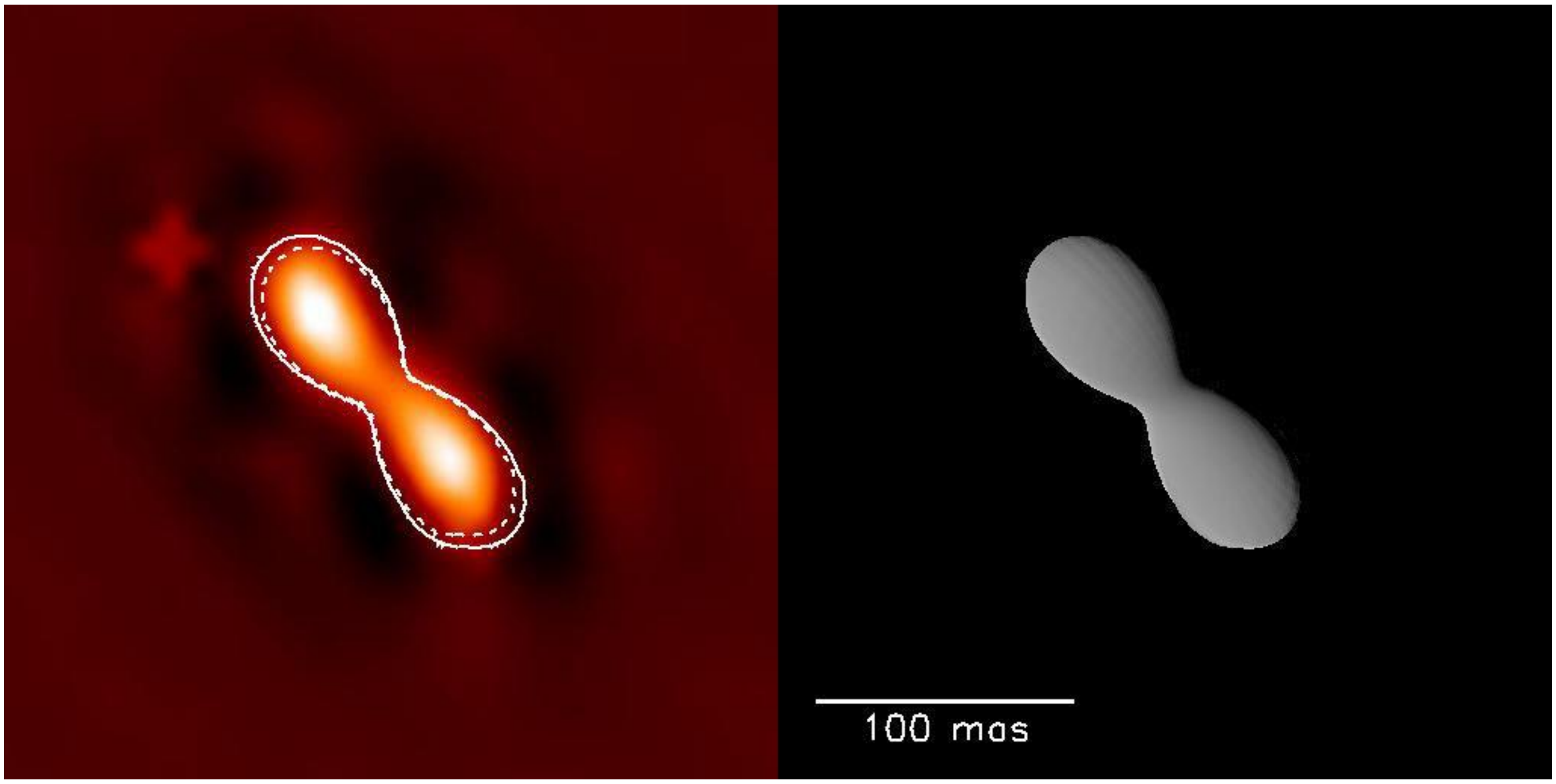

**Fig. 13 :** Stellar occultation of TYC 4909-00873-1 by Kleopatra on 2009 December 24th at 11:59 UTC. Kleopatra’s silhouette figures like Chinese shadows for each model with an equivalent radius adjusted to 62.5 km. Dumb-bell outline is about 250 km long and 70 km wide. Radar model outline is about 230 km long and slightly more than 80 km wide. The adaptive optics observations as well as the stellar occultation indicate that Kleopatra may be more of a dumb-bell than a dog-bone.

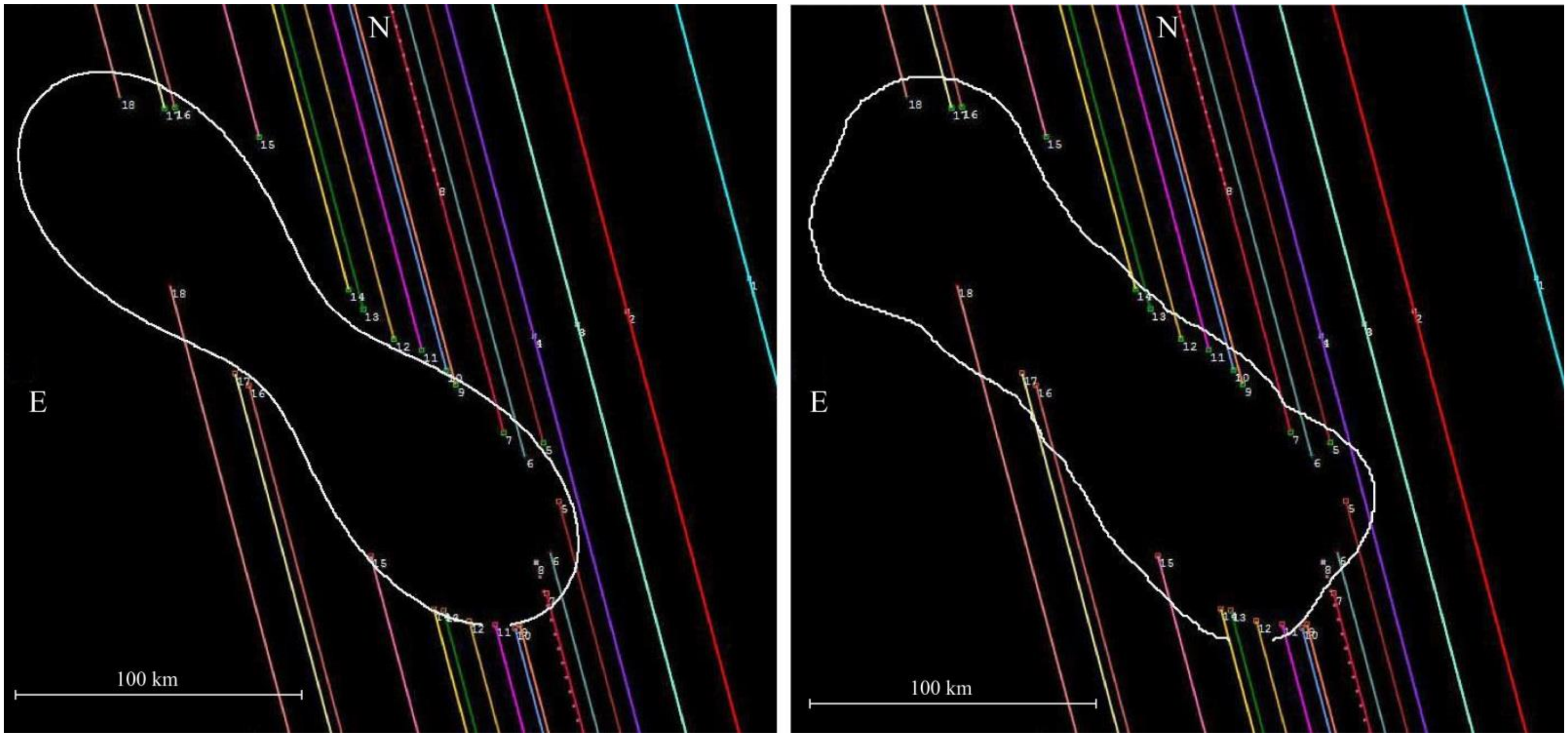

**Fig. 14:** Set of confidence regions for 624 Hektor derived from three light curves. See legend of Fig. 6 for the general description of the contours. Phase angle ($\alpha$) and aspect angle ($\psi$) are given for each geometry. The best solution (overlapping $1\sigma$ region in solid line) is obtained for $\Omega = 0.298 \pm 0.003$ and $k = 0.12 \pm 0.05$. With an assumed uncertainty of 0.05, 0.02 and 0.03 mag respectively in 1957, 1967 and 1968, this solution yields in 1957 $\chi^2 = 81.3$ for $\nu = 65$ degrees of freedom which gives a reduced $\chi^2$ of 1.32; in 1967, $\chi^2 = 102.7$ for $\nu = 65$ degrees of freedom which gives a reduced $\chi^2$ of 1.58; in 1968, $\chi^2 = 188.8$ for $\nu = 119$ degrees of freedom which gives a reduced $\chi^2$ of 1.58.

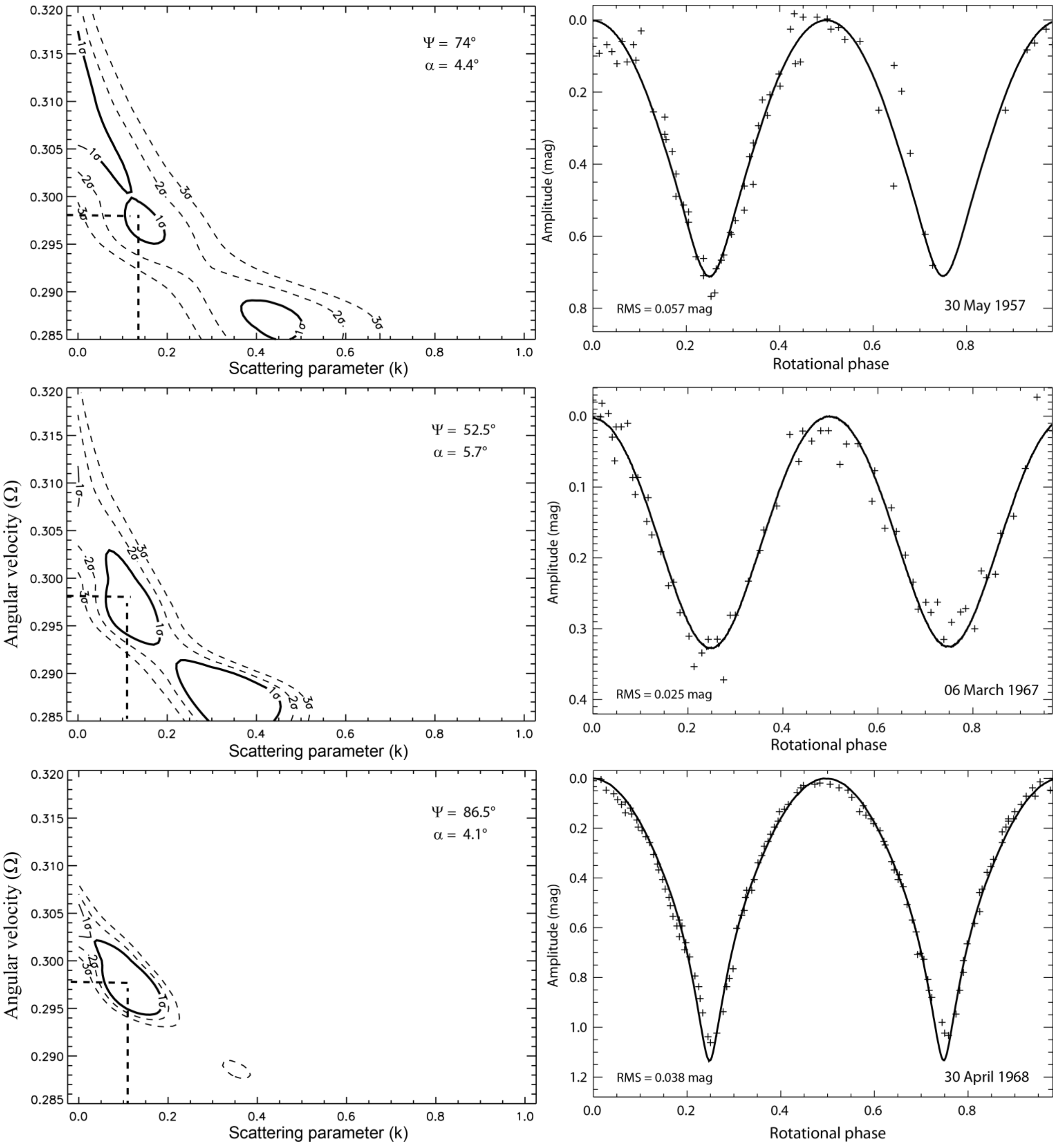

**Fig.15**: Adaptive optics image of Hektor taken with the 10-m Keck telescope on 16 July 2006 at 13:50 UTC. At this epoch, a scale of 100 mas represents 320 km. A Laplacian filter has been applied in order to enhance the overall contour of Hektor. The non-convex dumb-bell-shaped figure is shown as well. North is up and East is left. Contour lines of the model are superimposed to the observations for two values of the equivalent radius, 92 km (dotted line) and 112 km (solid line).

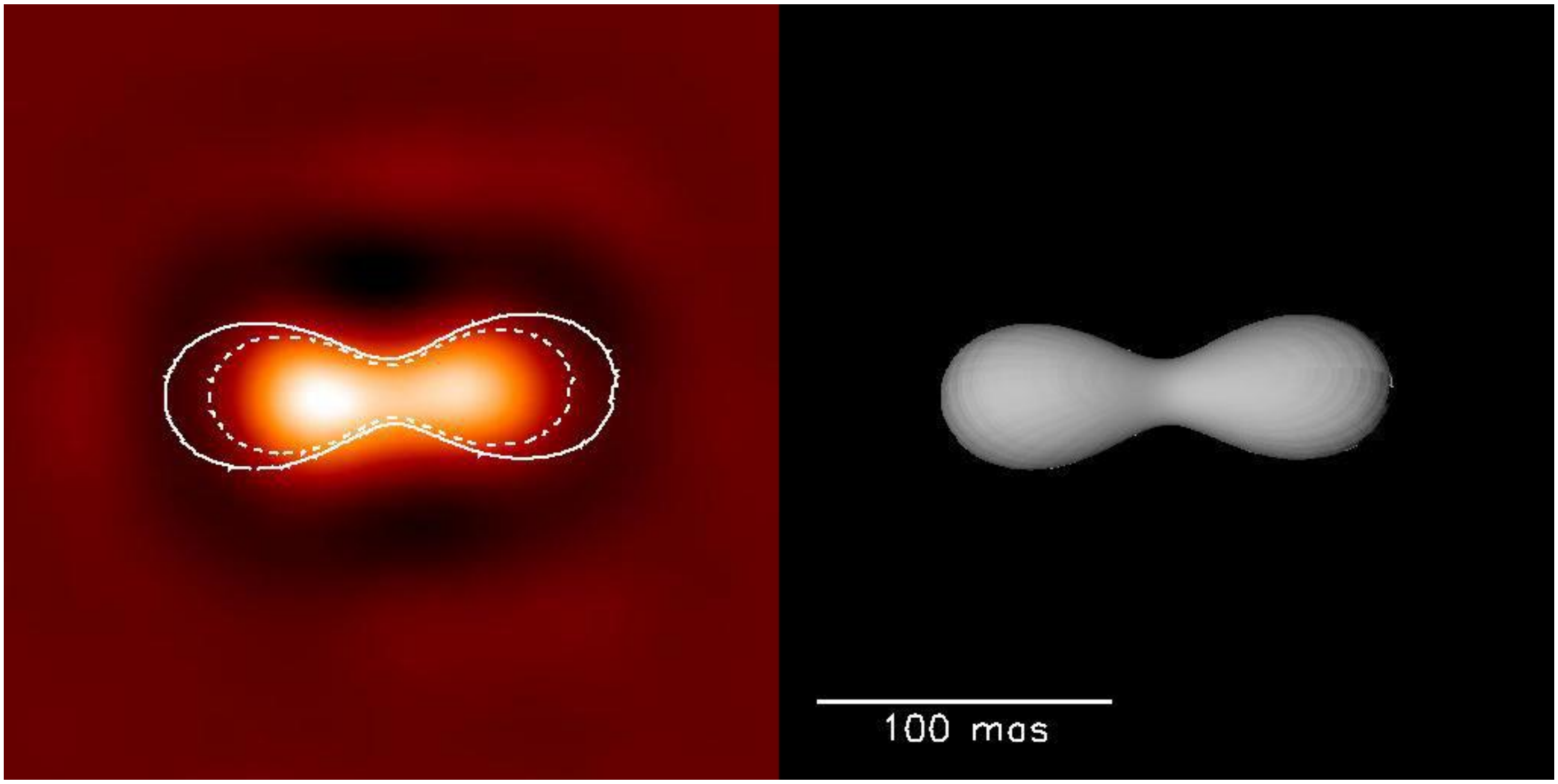

**Fig.A1**: Geometrical description of a cassinoide of parameter *e*. The cassinoide is computed and displayed for e = 1.04. The elongation of a cassinoide is given by $2\sqrt{1+e^2}/e^2$. Owing to its axisymmetry, the flattening is always zero. The angle $\theta_0$ defines the cassinoide parameter *e* such that $e = \sqrt{2\sin\theta_0}$. The limiting case is obtained with e = 1 which gives an elongation of $2\sqrt{2} = 2.83$ and a minimum value of $\theta_0 = 30°$.

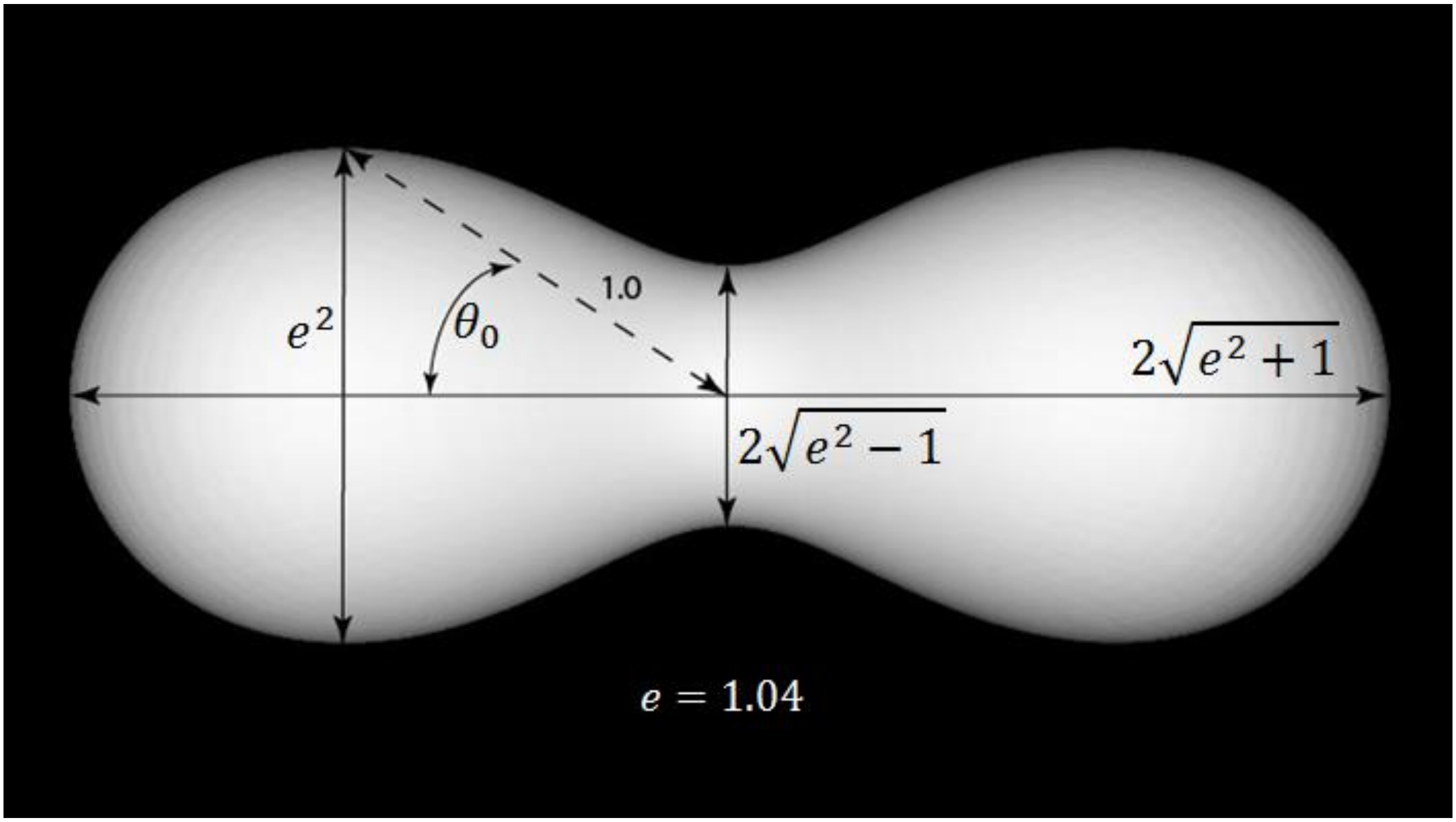